\documentclass[twocolumn,prc,amsmath,amssymb,aps,nofootinbib]{revtex4}
\usepackage{graphicx}
\usepackage{dcolumn}
\usepackage{bm}
\usepackage{soul}  
\usepackage[dvipsnames]{xcolor}
\usepackage{afterpage}
\usepackage[colorlinks=true]{hyperref}

\graphicspath{{figures/}}

\begin{document}

\author{Alice Bernard} 
\affiliation{CEA, DAM, DIF, 91297 Arpajon, France}
\affiliation{Université Paris-Saclay, CEA, LMCE, 91680 Bruyères-le-Châtel, France}

\author{David Regnier} 
\email{david.regnier@cea.fr}
\affiliation{CEA, DAM, DIF, 91297 Arpajon, France}
\affiliation{Université Paris-Saclay, CEA, LMCE, 91680 Bruyères-le-Châtel, France}

\author{Junah Newsome}
\affiliation{CEA, DAM, DIF, 91297 Arpajon, France}
\affiliation{Université Paris-Saclay, CEA, LMCE, 91680 Bruyères-le-Châtel, France}

\author{Paul Carpentier}
\affiliation{CEA, DAM, DIF, 91297 Arpajon, France}
\affiliation{Université Paris-Saclay, CEA, LMCE, 91680 Bruyères-le-Châtel, France}

\author{Noël Dubray}
\affiliation{CEA, DAM, DIF, 91297 Arpajon, France}
\affiliation{Université Paris-Saclay, CEA, LMCE, 91680 Bruyères-le-Châtel, France}

\author{Nathalie Pillet}
\affiliation{CEA, DAM, DIF, 91297 Arpajon, France}
\affiliation{Université Paris-Saclay, CEA, LMCE, 91680 Bruyères-le-Châtel, France}

\keywords{Monte Carlo, Markov chain, Nuclear fission, Bogoliubov vacuum}

\title{Probability distribution of observables from a Bogoliubov vacuum projected onto good particle number: application to scission configurations of an actinide}

\begin{abstract}

Nuclear fission dynamics described within nuclear energy density functional frameworks (EDF) have seen substantial advances in the last decade.
Part of this success stems from projection techniques, which allow the computation of probability distribution functions (pdf) for selected observables such as particle number and angular momentum of the fragments.
Predicting the pdf of other observables, such as the total kinetic energy of the fragments, remains undone.
This work proposes a method to determine the complete pdf of a new category of observables from a Bogoliubov vacuum projected onto good particle number.
It relies on sampling nucleonic configurations in coordinate and intrinsic-spin representation.
%
We assess the feasibility and convergence properties of the method and apply it to states representative of the scission of an actinide.
Fluctuations in fragment shapes, inter-fragment Coulomb and nuclear interaction as well as the corresponding torques are analyzed. 
We find that a significant fraction of the fluctuation of several measured fission observables is already present within the mean-field picture.
\end{abstract}

\maketitle

\section{Introduction}
\label{sec:intro}

The last decade has seen a wealth of progress in the theoretical description of the fission dynamics rooted in the EDF framework~\cite{schunck2016microscopic,schunck2022theory}.
Time-dependent mean-field calculations including pairing correlations allow for quantitative predictions of the energy balance and spin content
of the system after the generation of the primary fission fragments in the most populated fission mode~\cite{simenel2014formation,scamps2015superfluid,bulgac2016induced,bulgac2019fission,bulgac2020fissionfragment,bulgac2025timedependent,bjelcic2025excitation,bulgac2022angular,scamps2023spatial,scamps2024quantal,francheteau2024scission}.
The time-dependent generator coordinate method (TDGCM) based on a family of constrained Hartree-Fock-Bogoliubov states gives
qualitative estimations of the fragments mass and charge yields~\cite{regnier2016fission,regnier2019asymmetric,marevic2020fission,verriere2021microscopic,zhao2021microscopic,schunck2023microscopic,morfouace2025asymmetric}.
On top of that, the community is building new many-body methods aiming for a simultaneous description of the collective fluctuations and the dissipative aspects
of the fission process~\cite{DIETRICH2010249,bernard2011microscopic,tanimura2017microscopic,zhao2022timedependent,lau2022smoothing,marevic2023quantum,li2023generalizedb,lasseri2024generative,carpentier2024construction,li2024generalized,li2025fourier}.

In most of these approaches, the fissioning nucleus is depicted in terms of one or several Bogoliubov vacua
built with the neutrons and protons degrees of freedom.
We interpret the outcome of the dynamics in regards of the spatial one-body density of the Bogoliubov vacua involved;
especially its geometry discussed in terms of expectation values of multipole moment operators (\textit{i.e.}~\cite{scamps2018impact}).
Such a picture is intuitive and connects to the liquid-drop models at the origin
of our understanding of the fission process~\cite{meitner1939disintegration,bohr1939mechanism}.
However this simple description turns out to be insufficient to understand aspects of fission that stem from quantum fluctuations.
For instance, how do the axially symmetric one-body densities typically involved in fission simulations lead to rotating fragments ?
To answer this apparent dilemma one should consider the nucleus as a composite system made of randomly positioned nucleons.
The marginal pdf for the position of one nucleon may present an axial symmetry.
Yet many non-axial configurations for the cloud of nucleons will contribute to the wavefunction.
These configurations can lead to non vanishing fluctuations of observables such as the projection of the orbital momentum between the pre-fragments on an axis perpendicular to the fission axis.
Properly unraveling these effects not only requires computing expectation values of observables of interest but also the higher moments of their pdf.

In recent years, several studies successfully estimated the pdf of some fission observables
from a Bogoliubov vacuum based on projection techniques.
In Refs.~\cite{scamps2015superfluid,verriere2021microscopic,li2025microscopica}, the pdf of the mass and charge of the pre-fragments
were extracted close or after scission.
The same technique is at the heart of the determination of the angular momentum content after scission ~\cite{marevic2021angular,bulgac2022angular,scamps2023spatial,scamps2025uncertainty,marevic2025microscopic}.
In both cases, the observables under study are one-body observables known to be linked to a spontaneous symmetry breaking
of the EDF approach. This is what makes the computation of their pdf achievable in state-of-the-art calculations.
In contrast, similar predictions for kinetic energies of the fragments or for two-body observables such as the inter-fragment Coulomb repulsion were not achieved up to now.

In this work, we develop a method to study the pdf of a specific class of many-body observables from a Bogoliubov vacuum
projected on a good particle number.
To this end, we implement a sampler of the nucleons positions and intrinsic spin
consistent with the many-body density of the system.
The sampler enables estimating the pdf of any observable that is diagonal in the position/intrinsic spin representation.
In the context of fission, several observables of this kind are of particular interest, including the particle number, multipole moments, Coulomb interaction and any local two-body interaction.
In addition to quantifying fluctuations, the sampler may serve as a pedagogical tool for visualizing the nucleon cloud that constitutes the fissioning system.

In section \ref{sec:method} we detail the formalism required to sample the positions and intrinsic spins of the nucleons
and to estimate observables based on such a sample.
We study the feasibility and convergence of the method applied to light and heavy nuclei in section \ref{sec:numerics}.
Finally we study the pdf of some fission observables in two scission configurations of $^{252}$Cf in section~\ref{sec:physics}.
Conclusions and perspectives are given in section~\ref{conclusion}.

\section{Method}
\label{sec:method}

\subsection{Sampling configurations from a many-body state}

In the nuclear energy density functional framework the nucleons are point-like Fermions
that can be characterized in terms of a spatial coordinate $\bm{r}$ and the projection $\sigma$ of their intrinsic spin along the z-axis.
Following \cite{matsumoto2022visualization} we use the compact notation 
$r \equiv (\bm{r}\sigma) \equiv (x,y,z,\sigma)$ for the coordinates in the position/intrinsic spin representation
along with $c^{\dagger}_r$ and $c_r$ for the associated creation (resp. annihilation) operators.
With these notations, the Fock space for a gas of nucleons
is spanned by the family of position/intrinsic spin product states
\begin{equation}
  |r_1\cdots r_N\rangle = c^\dagger_{r_1} \cdots c^\dagger_{r_N} |0\rangle,
\end{equation}
where $|0\rangle$ is the particle vacuum.
For the purpose of our discussion, it is useful to define an ordering convention
in the position/intrinsic spin representation. 
This arbitrary ordering does not impact the results
and a possible choice is the lexicographic order on the $x,y,z,\sigma$ coordinates.
The family of ordered position/intrinsic spin product states 
\begin{equation}
    \left\{ |r_1\cdots r_N\rangle |\, r_1<\cdots<r_N \right\}
\end{equation}
form a basis of the Fock space which comes with the closure relation
\begin{equation}
     \int_{r_1<\cdots <r_N} 
     |r_1\cdots r_N\rangle \langle r_1\cdots r_N| dr_1
 \cdots dr_N = \hat{I} \label{eq:closure_ordered}
\end{equation}
Here we loosely used the integral symbol to designate integrals over the $x$, $y$ and $z$ axes of each one-body coordinate along with a sum over the intrinsic spins.
The probability to measure a gas of nucleons represented by a many-body state $|\psi\rangle$ in an ordered position/intrinsic spin product state reads
\begin{equation}
 p(|r_1\cdots r_N\rangle) = |\langle r_1\cdots r_N|  \psi \rangle|^2 .
\end{equation}

In the context of Monte Carlo sampling, it is convenient to work with a set of 
unordered configurations defined by n-tuple of 1-body coordinates
\begin{equation}
\left \{ (r_1 \cdots r_N) \in (\mathbb{R}^3\times \{\uparrow,\downarrow\})^N \right\}.
\end{equation}
We define a pdf on the set of configurations by
\begin{equation}
 p(r_1\cdots r_N) = \frac{1}{N!} \, |\langle r_1\cdots r_N|  \psi \rangle|^2 .
 \label{eq:pdf}
\end{equation}
Note that the probabilities are invariant with permutations of the coordinates.
Owing to the closure relation~\eqref{eq:closure_ordered}, this pdf is correctly normalized in the space of configurations
\begin{equation}
      \int_{r_1,r_2 \cdots ,r_N} 
     p(r_1\cdots r_N)
dr_1 \cdots dr_N = 1.
\end{equation}
The class of configurations having the same coordinates up to a permutation
is representative of one ordered position/intrinsic spin product state 
and each configuration within this class has the same statistical weight.
Sampling configurations from this pdf gives an insight into the content of the many-body wavefunction in the position/intrinsic spin representation and
will also enable us to compute observable moments in a Monte Carlo way.

To generate a sample of configurations from the pdf~\eqref{eq:pdf} we rely on Markov chains~\cite{fishman1996monte}.
Markov chains are particularly suited to tackle the large dimensionality of the configuration space (more that 600 for an actinide).
In addition, only ratios of probabilities need to be estimated which enables us to avoid
computing costly normalization factors that may appear in \eqref{eq:pdf}.
In this work we build Markov chains in the space of configurations
out of the simple Metropolis iterative algorithm.
One step of the Metropolis algorithm can be summarized as:
\begin{enumerate}
\item From a configuration $(r_1^i\cdots r_N^i)$ at the i'th iteration of the chain,
sample a candidate configuration $(\tilde{r}_1^{i}\cdots \tilde{r}_N^{i})$ for the next iteration.
\item Compute the ratio of probability 
$\alpha = \frac{p \left( \tilde{r}_1^i \cdots \tilde{r}_N^i \right)}{p\left( r_1^i \cdots r_N^i \right)}$
\item Sample a random number $\xi$ in a uniform distribution spanning $[0,1]$.
\item Choose the candidate configuration $(\tilde{r}_1^i \cdots \tilde{r}_N^i)$ for the next iteration if $\alpha > \xi$. Otherwise stick to $(r_1^i\cdots r_N^i)$ for the next iteration.
\end{enumerate}
The candidate configuration is sampled by perturbing one nucleon's coordinate from the i'th configuration.
Changing only one nucleon's coordinates per iteration ensures a non vanishing acceptance rate of 
the Metropolis step.
In details, we
\begin{enumerate}
\item sample the index of the particle that will be perturbed from a discrete uniform distribution,
\item sample and apply a small perturbation on the $x,y$ and $z$ coordinates of this particle from independent normal distributions 
characterized by a spatial width $\sigma_{space}$,
\item randomly flip the intrinsic spin projection of the particle with a probability $P_{flip}$.
\end{enumerate}
The width of the space perturbation along with the spin flip
probability are typically chosen as 1 fm and 10\% (cf. section \ref{sec:numerics}).

The Metropolis algorithm guarantees that, after a sufficient number of iterations, the pdf of the sampler corresponds to the target pdf $p(r_1\cdots r_N)$.
To extract a representative sample, a standard procedure consists in warming up the Markov chain during
$N_{burn-in}$ iterations called the burn-in period.
Only after this burn-in period, one starts to record
configurations from the sampler.
In addition, events generated in successive iterations of the Markov chain are strongly correlated.
To mitigate this effect and obtain a final sample 
representative of independent random draws from the target distribution,
we only retain one configuration every $N_{jump}$ iterations.
If the burn-in and jump are properly taken into account, 
the Metropolis algorithm can generate an
arbitrary number of configurations representative of successive independent random draws
from the target distribution.
How we define the burn-in period, the jump as well as the spatial width and spin flip probabilities
to ensure a proper convergence and minimize the numerical cost will be discussed in section \ref{sec:numerics}.

\subsection{Estimating observables}

As widely used in the context of variational quantum Monte Carlo~\cite{lynn2019quantum}, a configuration sampler enables the computation of 
expectation values of observables.
In this work, we restrict ourselves to observables $\hat{O}$ that are diagonal in the 
position/intrinsic spin representation.
In other words, for any pair of position/intrinsic spin product states:
\begin{equation}
 \langle r_1 \cdots r_N| \hat{O} | r_ 1'\cdots r_N'\rangle  = 
 O(r_1\cdots r_N)
 \delta_{r_1 r'_1} \cdots \delta_{r_N r'_N}
\end{equation}
The kernel $O(r_1 \cdots r_N)$ is invariant by permutations of the coordinates
and matches the diagonal matrix elements of $\hat{O}$ in the basis of the ordered position/intrinsic spin product states.
Although this property restricts the domain of applicability of our method, many observables encountered in 
fission studies possess this property.
On top of that, it enables a straightforward computation of the whole pdf of the observable, 
without the appearance of the so called sign problem~\cite{lynn2019quantum}.
Using the closure relation \eqref{eq:closure_ordered} along with the diagonal nature of the observable,
it can be shown that the expectation value of a diagonal observable reduces to an integral over the space of configurations:
\begin{align}
 \langle \hat{O} \rangle &= 
 \int_{r_1\cdots r_N} O(r_1\cdots r_N) \,
 p(r_1\cdots r_N) dr_1\cdots dr_N
\end{align}
This is a many-dimensional integral with a product integrand well suited for a
Monte Carlo estimation.
A sample of $L$ configurations independently drawn from the distribution
$p(r_1\cdots r_N)$ gives an unbiased estimator of the expectation value $\mu_{ \langle \hat{O} \rangle }$:
\begin{equation}
 \mu_{ \langle \hat{O} \rangle } =  \frac{1}{L} \sum_{l=1}^L O( r_1^l \cdots r_N^l ),
\end{equation}
along with an estimator of the corresponding statistical uncertainty
\begin{equation}
 \sigma_{stat}\left(\mu_{ \langle \hat{O} \rangle }\right) =  
 \left[
  \sum_{l=1}^L 
  \frac{
 \left( 
 O( r_1^l \cdots r_N^l ) - \mu_{ \langle \hat{O} \rangle }
 \right)^2}
 {L(L-1)}
 \right]^{-\frac{1}{2}}.
 \label{eq:sigma_stat}
\end{equation}
To summarize, the random variable $O(r_1\cdots r_N)$ obtained by sampling configurations
has $\langle \hat{O} \rangle$ for expectation value.

With diagonal observables, we can even go one step further and remark that the pdf of the random variable $O(r_1\cdots r_N)$ matches the pdf of the measurements of the observable $\hat{O}$.
One possible justification comes from the fact that $\hat{O}$ elevated to some
power is still diagonal in the position/intrinsic spin representation. Hence, all the moments of 
its pdf equal the moments of the random variable $O(r_1\cdots r_N)$.
We end up with a straightforward method to estimate the pdf of observables: (i) sample configurations, (ii) compute the observable kernel for each configuration (iii) infer the pdf of the observable from this representative sample.

In this work, we focus on one-body and two-body observables diagonal in the position/intrinsic spin
representation
\begin{align}
 \hat{T} &= \int_r t(r) \, c^{\dagger}_{r} c_{r}, \\   
 \hat{V} &= \frac{1}{2}\int_{r r'} 
 v(r, r')
 c^{\dagger}_{r} \, c^{\dagger}_{r'} c_{r'} c_{r}.
\end{align}
The corresponding kernels can be cast as permutation invariant 
sums over nucleons (resp. pairs of nucleons).
\begin{align}
  T(r_1 \cdots r_N) &= \sum_{i=1}^{N} t(r_i), \\
  V(r_1 \cdots r_N) &= 
  \frac{1}{2} \sum_{i<j=1}^{N} \left[ v(r_i, r_j) + v(r_j,r_i) \right].
  \label{eq:2bdykernel}
\end{align}

\subsection{Probabilities for Bogoliubov vacua projected on good particle number}

In this study we describe atomic nuclei with many-body states $|\psi\rangle$ being
tensor products of a neutron and proton gases.
\begin{equation}
    |\psi\rangle = |\psi_n\rangle \otimes |\psi_p\rangle.
    \label{eq:tensor_state}
\end{equation}
One configuration of the system is given by a configuration of the N neutrons
and a configuration of the Z protons.
The pdf for the configurations is the product of the neutron and proton independent pdfs.
The neutron (resp. proton) wavefunctions considered here are fully paired Bogoliubov vacua projected
onto good particle numbers.
We can express for instance the neutron part in its canonical basis as 
\begin{equation}
    |\psi_{n}\rangle = \hat{P}_N \prod_{k=1}^{K} (u_k + v_k a^\dagger_k a^\dagger_{\bar{k}}) |0\rangle.
    \label{eq:wavefunction}
\end{equation}
where $\hat{P}_N$ is the projector on the neutron number $N$, $a_k^\dagger,a_{\bar{k}}^\dagger$ stand for the creators of the single-particle states of the canonical basis and $u_k, v_k$ are real numbers verifying $u_k\neq 0$ and $u_k^2 + v_k^2=1$.

To iterate Markov chains, we need to build and evaluate a function proportional to the 
configuration pdf.
Such a function was already used for instance in Ref.~\cite{matsumoto2022visualization} and we only recall here the necessary formulas.
For the wavefunction \eqref{eq:wavefunction}, 
the square modulus of interest can be cast into the form of a determinant
\begin{equation}
    |\langle r_1,\cdots,r_N| \psi_n \rangle|^2 
    \propto
    |det(\mathcal{Z})|.
    \label{eq:probability_amplitude}
\end{equation}
The skew-symmetric $\mathcal{Z}$ matrix contains information about pairs of particles
\begin{equation}
\forall i,j:
\mathcal{Z}_{ij} = 
Z(r_i,r_j) - Z(r_j,r_i),
\end{equation}
with
\begin{equation}
 Z(r,r') = \sum_{k=1}^{K} \frac{v_k}{u_k} \varphi_k(r)\varphi_{\bar{k}}(r').
 \label{eq:Zrrp}
\end{equation}
Here we used $\varphi_k(r)$ to designate the canonical single-particle wavefunctions
evaluated at the coordinate $r$.
Note that extensions of this formula to odd or incompletely paired systems are possible~\cite{matsumoto2022visualization} but were not required in the present study.

To conclude, we implemented the whole Markov Chain Monte Carlo (MCMC) method described in this section
into a C++ code. We released this implementation as an open-source
code named {\tt NucleoScope}~\cite{regnier2026nucleoscope}.

\section{Numerics}
\label{sec:numerics}

\begin{figure}
\includegraphics[width=0.50\textwidth]{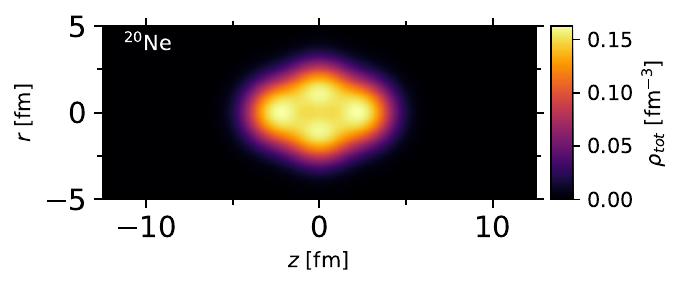}
\includegraphics[width=0.50\textwidth]{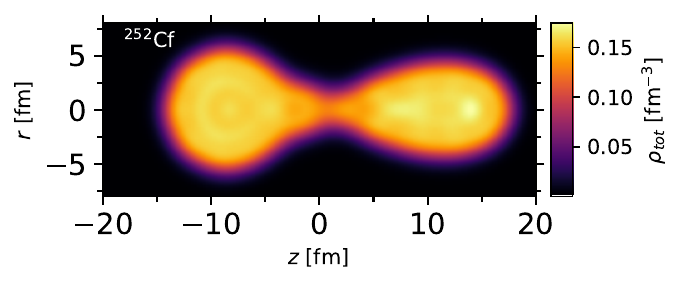}
\caption{Local nucleon one-body density for the $^{20}$Ne (top) and the $^{252}$Cf (bottom) Bogoliubov vacua considered in Sec.~\ref{sec:numerics}.}
\label{fig:numerics_densities}
\end{figure}

In this section, we investigate the numerical aspects of our method to estimate observable pdfs.
Because the dimensionality of the configuration space increases linearly with the number of particles, we perform numerical tests for two different systems.
We first sample the nucleon coordinates from a wavefunction representing the ground state of $^{20}$Ne, a light 
open-shell nucleus.
We then undertake the same work for a $^{252}$Cf nucleus close to its most favored scission configuration.
In both cases, the wavefunction is a Bogoliubov vacuum projected onto good particle number.
The Bogoliubov vacua were obtained by solving constrained Hartree-Fock-Bogoliubov equations with a Gogny D1S energy density functional and the code {\tt HFB3}~\cite{dubray2025hfb3}.
The one-body densities of the Bogoliubov states are shown in FIG.~\ref{fig:numerics_densities}.
We provide the details of these calculations in the section~\ref{sec:wavefunction_details}.

\subsection{State discretization}
\label{sec:state_discretization}

To compute the probability amplitude for any configuration 
with \eqref{eq:probability_amplitude},
we first generate and store the canonical representation of the Bogoliubov vacua under study. 
To speed up our evaluation, we truncate this representation keeping only the most
populated canonical states. 
We use a truncation threshold corresponding to a maximal error of $10^{-4}$ particles on the Bogoliubov vacua.
For the neutron part it translates into
\begin{equation}
\sum_{k \in {\mathcal{T}}} v_k^2 < 10^{-4} \, N,
\end{equation}
where $\mathcal{T}$ is the truncated set of canonical states.
The effect of this truncation on multipole moments of the Bogoliubov vacua projected onto good particle number was found negligible compared to other sources of numerical bias investigated in this work.

The canonical states are spatially discretized on a Lagrange mesh introducing a box size and a step parameter in each direction~\cite{baye2015lagrangemesh}.
For the $^{20}$Ne, we use a cubic box with dimensions $L_x\times L_x \times L_x$ and a regular mesh with cubic cells of volume $dx^3$.
For elongated $^{252}$Cf states we use a rectangular box with dimensions $L_x \times L_x \times 2L_x$.
As detailed in appendix~\ref{sec:wavefunction_details}, we chose a box size 
$L_x=32$ fm that guarantees the convergence of the first two non-vanishing multipole moments of the projected Bogoliubov vacua for both the light and heavy nuclei.
To accelerate our calculation we approximate the canonical wavefunctions to constants
inside each mesh cell during the evaluations of Eq.~\eqref{eq:Zrrp}.
To mitigate the bias coming from this crude interpolation scheme, we
use a small cell size of $dx=0.5$ fm for our physical studies.
We provide an idea of the error coming from such a spatial discretization in the benchmark of Sec.~\ref{sec:benchmark}.

\subsection{Markov Chains}

\subsubsection{Initialization}

Starting a Markov chain requires an initial configuration. 
To avoid large burn-in periods we try building initial configurations with a reasonably high probability.
We start by computing the total spatial one-body density of the state
and extract the mesh cells that have a density above 2.5\% of the saturation density (0.16 fm$^{-3}$).
For each nucleon, we choose randomly one of these cells and place it at its center.
One nucleon out of two is assigned a spin up while the other starts with a spin down.
This simple method dispatches the nucleons within the nuclear volume while respecting the
Pauli principle.

\subsubsection{Burn-in period}

In MCMC, the sampled configurations begin to be representative of the target pdf only after
a certain number of iterations~\cite{fishman1996monte}. These first $N_{burn-in}$ iterations form the burn-in period and should be discarded from the final sample.
To evaluate the necessary length of the burn-in period, we rely on a statistical criteria proposed by Brook and Gelman~\cite{brooks1998general,gelman1992inference}.
This Gelman-Rubin ratio is estimated on the last $n$ iterations of several independent Markov chains of length $2n$ as detailed in \ref{sec:gelman-rubin}. 
It tends to one when the Markov chains reach their stationary regime.
Because of the large dimensionality of the configuration space, we apply this criteria
not on the configurations themselves but separately on each observable under study.
Some observables may require longer burn-in than others and we pragmatically 
choose the largest burn-in needed for the set of observables under study.

\begin{figure}[h!]
    \centering
    \includegraphics[width=1.0\linewidth]{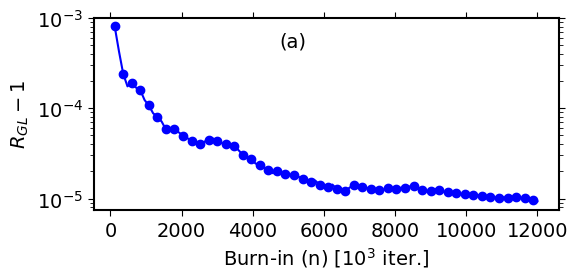}
    \includegraphics[width=1.0\linewidth]{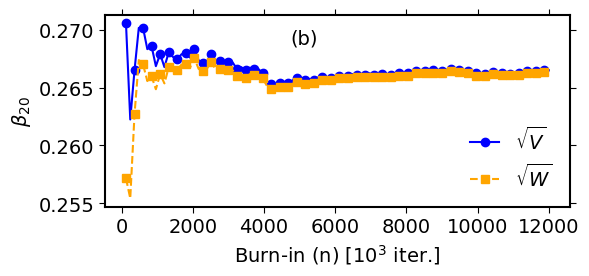}
    \caption{(a) Gelman-Rubin ratio minus one obtained for $\beta_{20}$ in the $^{20}$Ne ground state. (b) Square root of the estimators of the variance of $\beta_{20}$ from the Markov chains sample ($V$) and of the intra-chain variance ($W$).}
    \label{fig:r}
\end{figure}

Figure~\ref{fig:r} shows the results obtained for the $\hat{\beta}_{20}$ observable as a function of $n$
in the case of $^{20}$Ne. These calculations were obtained with a spin flip probability of $P_{flip}=0.1$ and spatial width $\sigma_{space}=1$ fm.
The Gelman-Rubin ratio converges rapidly to one from above. It becomes lower than 1.005
around $n=6.10^4$. 
At the same time, the average intra-chain variance $W$ seems to be stabilized at percent level.
Similar burn-in periods were obtained for a two-body observable representing the Coulomb interaction indicating that $n=10^5$ is the correct order of magnitude for burn-in in light nuclei.
For $^{252}$Cf, we find that $n=10^6$ is a conservative order of magnitude.

\subsubsection{Choice of $\sigma_{space}$ and $P_{flip}$}
\label{sec:choice_mc_param}

The burn-in period may vary as a function of the pdf of the perturbation applied at each iteration.
In this section, we search for spatial width and spin flip probabilities that 
ensure a good compromise between short burn-in periods and large acceptance rate.
We seek for typical acceptance rate of 0.5 to avoid the situation were a configuration stays
identical during a large number of iterations, that would ultimately lead to large
autocorrelation in the final sample.
To do so, we performed series of calculations varying $\sigma_{space}$ and $P_{flip}$.
For each calculation, we estimate the burn-in period as the shortest $n$ for which:
\begin{enumerate}
 \item the Gelman-Rubin ratio $R_{GR}(n) < 1.005$,
 \item the average intra-chain standard deviation is stabilized within 1\%, \textit{i.e.}
 $$
 max_{\left\{n'>n\right\} }
 \frac{|\sqrt{W(n')} - \sqrt{W(n_{max}})|}
 {\sqrt{W(n_{max})}} < 0.01.
 $$
\end{enumerate}
We report in \ref{sec:parameter_impact_on_burnin} the detailed results.
Overall, we find that the burn-in period does not vary drastically as a function of these parameters. All our calculations give results within one order of magnitude. 
As a consequence, we stick to our guessed probability $P_{flip}=0.1$ for the rest of this work.
We also choose $\sigma_{space} =1$ fm that typically gives an acceptance rate close to 0.5.

\subsubsection{Mitigation of autocorrelation}

A drawback of Markov chain samplers is the high correlation that may be present between successive configurations.
Such a correlation may bias estimators that are built assuming independent draws from the target
pdf as for instance the statistical uncertainty estimators~\eqref{eq:sigma_stat}.
In this work, we estimate the auto-correlation of a sample of an observable $\{O_i |  \in[1,n] \}$ as
\begin{equation}
	c = \frac{ cov(O_i, O_{i+1})}{\sigma(O_i) \sigma(O_{i+1})},
\end{equation}
where $\sigma(O_i)$ (resp. $\sigma(O_{i+1})$) is the standard deviation of the first (resp. last) $n-1$ events of the sample
and $cov(O_i, O_{i+1})$ is the covariance between the $n-1$ pairs of successive events.
With the Markov chains built in section~\ref{sec:method}, we find that all the observables possess an autocorrelation
above 96\% for $^{20}$Ne and above 99\% for $^{252}$Cf.

\begin{figure}[!h]
    \centering
    \includegraphics[width=1.0\linewidth]{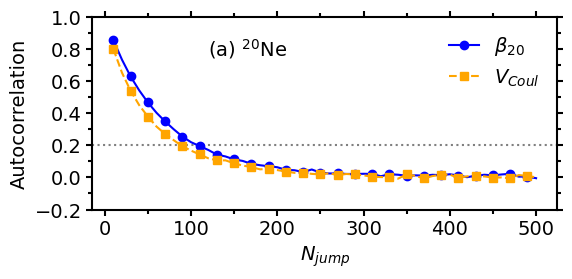}
    \includegraphics[width=1.0\linewidth]{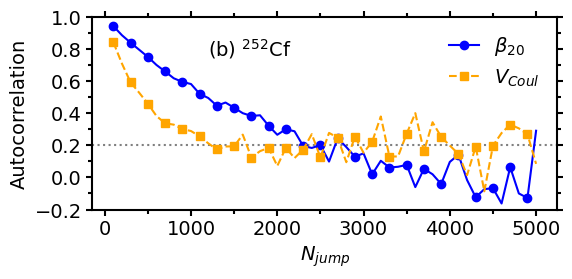}
    \caption{(a) Autocorrelation between successively recorded configurations from one Markov chain.
    We focus on the quadrupole moment and the Coulomb interaction in a $^{20}$Ne ground state. (b) Idem for a pre-scission configuration of $^{252}$Cf.}
    \label{fig:auto}
\end{figure}

To mitigate this issue, a standard procedure consists in recording only one draw every $N_{jump}$.
Large $N_{jump}$ values reduce the autocorrelation at the price of decreasing the number of 
recorded configurations.
We show in FIG.\ref{fig:auto} the evolution of the autocorrelations as a function of the jump
for the elongation $\hat{\beta}_{20}$ and the Coulomb interaction $\hat{V}_{Coul}$.
These results were obtained from one Markov chain containing $4.10^6$ iterations after a burn-in
period of $10^6$ iterations.
We see that the autocorrelation decreases monotonously with the jump parameter.
The fluctuations in our estimation of the autocorrelation for large values of the jump in $^{252}$Cf 
are likely coming from a lack of statistics ($800$ events only to estimate the point $N_{jump}$=5000).
Already with these results, we see that reaching a target autocorrelation of 0.2 \cite{koonin1997shell} typically requires 
$N_{jump}\simeq100$ for $^{20}$Ne and $N_{jump}\simeq 3000$ for $^{252}$Cf.
A scaling of the required jump with the number of particles is expected as one iteration only 
modifies one proton and one neutron.
In the following, we will work with a jump of 2000 when studying the fission of $^{252}$Cf.

\subsection{Benchmark with deterministic calculations}
\label{sec:benchmark}

This section aims at giving a final assessment on the convergence of the Monte Carlo
approach to estimate observables.
We compare MCMC results with deterministic calculations for
the first moments of several one-body observables 
and for the expectation value of a two-body observable.
The deterministic calculations were directly performed in the harmonic oscillator basis without
errors coming from changing the spatial discretization (see. \ref{sec:state_discretization}).

For this benchmark, we recorded $10^6$ events out of 10 independent Markov chains
with a burn-in period of $10^5$ and a jump of 100 for $^{20}$Ne (resp. a burn-in period of $10^6$ and a jump of 2000 for $^{252}$Cf).
TABLES \ref{tab:bench_20Ne} and \ref{tab:bench_252Cf} summarize the results
obtained with two spatial cell sizes.
%
\begin{table}[h!]
\begin{tabular}{l|ll|l}
\hline
\hline
  Method                  &  MCMC      &  MCMC        & Det.  \\
  dx                      & 1.     & 0.5              &   \\
\hline
$\langle \beta_{20} \rangle$  & 0.565 $\pm 8.10^{-4}$   & 0.565 $\pm 8.10^{-4}$   & 0.565 \\
$\langle \beta_{40} \rangle $ & 0.573 $\pm 2.10^{-3}$   & 0.573 $\pm2.10^{-3}$    & 0.573 \\
$\langle V_{Coul}  \rangle$ [MeV] & 19.44 $\pm1.10^{-2}$ & 19.57 $\pm1.10^{-2}$   & 19.88 \\
$ \sigma( \beta_{20} )$       & 13.89             & 13.50              & 13.37 \\
$ \sigma( \beta_{40} )$       & 0.775             & 0.743              & 0.733 \\
$ \sigma( V_{Coul})$ [MeV]      & 2.243             & 2.201              &  \\
\hline
\hline
\end{tabular}
\caption{Benchmark of several observables of $^{20}$Ne computed from the MCMC and in a deterministic way (Det.) using projection techniques directly in the harmonic oscillator basis.
The error bars provided for expectation values correspond to three times the estimator of the 
Monte Carlo statistical uncertainty (99.7\% confidence intervals).}
\label{tab:bench_20Ne}
\end{table}
%

%
\begin{table}[h!]
\begin{tabular}{l|ll|l}
\hline
\hline
  Method                  &  MCMC      &  MCMC        & Det.  \\
  dx                      & 1.     & 0.5              &   \\
\hline
$\langle \beta_{20} \rangle$      & 4.108  $\pm 3.10^{-4}$   & 4.108 $\pm 3.10^{-4}$   & 4.108  \\
$\langle \beta_{30} \rangle $     & 2.136  $\pm 8.10^{-4}$   & 2.138 $\pm8.10^{-4}$    & 2.134  \\
$\langle V_{Coul}  \rangle$ [MeV] & 844.94 $\pm5.10^{-2}$    & 846.45 $\pm5.10^{-2}$   & 846.94 \\
$ \sigma( \beta_{20} )$           & 0.099                    & 0.098                   & 0.107 \\
$ \sigma( \beta_{30} )$           & 0.269                    & 0.266                   & 0.257 \\
$ \sigma( V_{Coul})$ [MeV]        & 2.243                    & 2.201              &  \\
\hline
\hline
\end{tabular}
\caption{Same as Tab.~\ref{tab:bench_20Ne} for $^{252}$Cf.}
\label{tab:bench_252Cf}
\end{table}
%

Overall, this benchmark shows a good agreement between the results obtained
from the two approaches.
Expectation values of multipole moments are reproduced within the statistical 
uncertainty inherent to the sample size.
For the Coulomb energy, we note errors of the order of a few percents that seem to reduce with the cell size. 
This is likely 
connected to the constant wavefunction approximation in each cell coupled with the Pauli principle that prevents two nucleons of the same spin and isospin to lie in the same cell.
This effect removes some configurations with high Coulomb energies.
It is consistent with the fact that the MCMC calculations converge to the deterministic
value from below.
Standard deviations of the observables under study are converged within a few percents.
The size of the cell plays a significant 
role in the remaining error, hence our final choice of dx $=0.5$ fm for the cells dimension.
If sub-percent precision would be required, future studies may correct this error by introducing a better interpolation scheme for the canonical wavefunctions.

\section{Static properties of Bogoliubov vacua close to scission}
\label{sec:physics}

In this section, we sample the position and intrinsic spin of the nucleons for two
projected Bogoliubov states representing a $^{252}$Cf nucleus close to scission.
The generated sample enables us to investigate the probability distribution of
various observables in this crucial step of fission.
Due to its spontaneous fission, this nucleus is a standard for fission studies
with high statistic measurements of many fission observables (\textit{e.g.}~\cite{gook2014prompt}).

\subsection{Geometry of the scission configurations}

\begin{figure}[!h]
    \centering
    \includegraphics[width=1.0\linewidth]{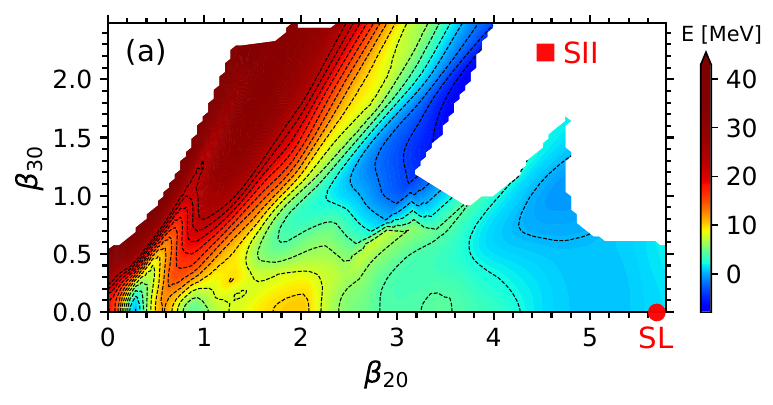}
    \includegraphics[width=1.0\linewidth]{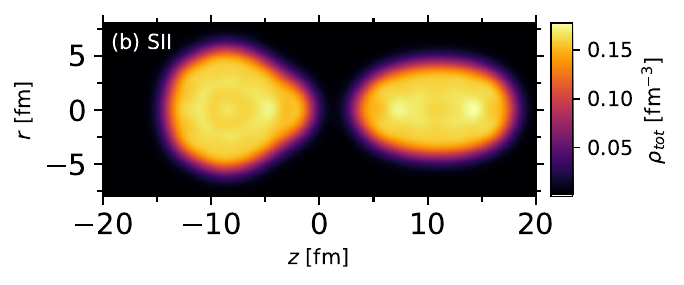}
    \includegraphics[width=1.0\linewidth]{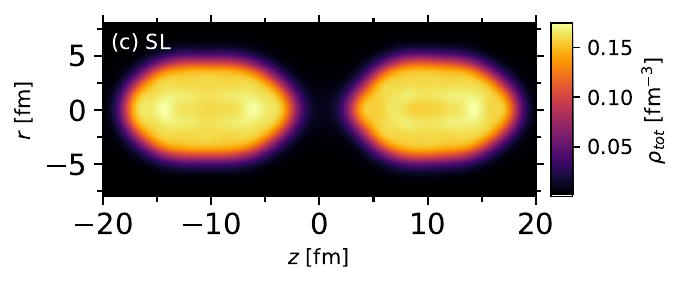}
    \caption{(a) Energy landscape of $^{252}$Cf obtained with constraints on the quadrupole and octupole moments. 
    It corresponds to the HFB energy with a vibrational zero-point-energy correction.
    The energy is relative to the ground state.
    Calculations were limited to configurations with a number of particle in the neck $Q_{neck}>5$.
    (b) Local one-body density of a Bogoliubov vacuum representative of the SII fission mode.
    (c) Local one-body density of a Bogoliubov vacuum representative of the SL fission mode.}
    \label{fig:pes}
\end{figure}

A prominent mode in the fission of actinides is the standard II (SII)~\cite{brosa1990nuclear}.
This early branch of the fission output channels is characterized by an asymmetric repartition of matter
between the heavy fragment, close to magicity, and an elongated light fragment.
In the context of constrained Hartree-Fock-Bogoliubov (HFB) calculations, the SII mode 
manifests itself by a valley in the quadrupole-octupole HFB energy surface.
We show in FIG.~\ref{fig:pes} such an energy surface obtained for the $^{252}$Cf
with the Gogny D1M effective interaction~\cite{goriely2009first}.
The corresponding constrained HFB states all possess a significant number of particle remaining in the neck.
This translates into an isoscalar neck operator expectation values $\langle \hat{Q}_{neck}\rangle >5$
along with a residual inter-fragment nuclear interaction
of the order of a hundred MeV~\cite{younes2011nuclear,carpentier2024microscopic}.
For the sake of completness we recall here the definition of the neck operator
\begin{align}
 \hat{Q}_{neck} = exp\left(- \frac{(z-z_{neck})^2}{a_{neck}} \right) c^{\dagger}_r c_r.
\end{align}
We use the standard convention $a_{neck}=1$ fm$^2$.
The neck position
$z_{neck}$ is chosen as the z-coordinate minimizing the local one-body density.

In this study we want to investigate a nuclear configuration representative of the SII mode in a regime where the nuclear inter-fragment interaction becomes significantly smaller than the Coulomb repulsion.
To do so we solve the constrained HFB equations with constraints on $\beta_{20},\beta_{30}$ and the neck operator $\langle \hat{Q}_{neck}\rangle = 0.4$.
With this choice of number of particles in the neck, we expect an inter-fragment nuclear interaction
of the order of 20 MeV~\cite{younes2011nuclear,carpentier2024microscopic} along with Coulomb repulsion of the order
of 180 MeV.
On top of that, the constraint on $\beta_{30}$ was roughly tuned so to stick to a number of particle close to 142 in the heavy fragment.
We project the obtained Bogoliubov vacuum onto good particle number to produce the input wavefunction
to our Monte Carlo sampler.
We carry a similar procedure to generate a wavefunction representative of the super-long (SL) mode.
The deformation of the two Bogoliubov vacua used are marked with red symbols in the panel (a) of FIG.~\ref{fig:pes},
while we plot their local one-body density in the panels (b) and (c).
%

For the two Bogoliubov vacua under study, we sampled the position and intrinsic spin of the
nucleons following the method detailed in~\ref{sec:method}.
A total of 256000 configurations were recorded out of 256 independent Markov chains.
Each Markov chain uses a jump of 2000 iterations and a burn-in period of $2.10^6$ iterations.
We extract from this sample numerous distributions of one-body observables and summarize
their expectation values and standard deviations in TABLE~\ref{tab:observables1}.
Fragment observables were obtained by introducing a Heaviside function $\Theta_{z>z_{neck}}$
into the observable kernels. This method introduced in Ref.~\cite{simenel2010particle} allows to only account for the nuclear matter present in the right or left half-space.
 
%
\begin{table}[!h]
\begin{tabular}{l|ll|ll}
\hline
\hline
 & SII & & SL & \\
 & $\langle\hat{O}\rangle$ & $\sigma(\hat{O})$ & $\langle\hat{O}\rangle$ & $\sigma(\hat{O})$ \\
  \hline
$x_{CM}$ [fm] & 0. & 0.09       & 0.   & 0.13 \\
$z_{CM}$ [fm] & 0. & 0.15       & 0.   & 0.29 \\
$\beta_{20}$ & 4.28 & 0.10      & 5.35 & 0.12 \\
$\beta_{30}$ & 2.29 & 0.28      & 0.01 & 0.37 \\
$Q_{neck}$   & 0.39 & 0.43      & 0.40 & 0.48 \\
\hline
$A_H$ & 143.8 & 1.45            & 127.0 & 2.14 \\ 
$Z_H$ & 55.6  & 1.06            & 49.2  & 1.83 \\ 
$x^H_{CM}$ [fm] & 0. & 0.13     & 0.    & 0.13 \\
$z^H_{CM}$ [fm] & -8.14 & 0.14  & -10.3 & 0.17 \\
$\beta_{20}^H$  & 0.17 & 0.05   & 0.68  & 0.07 \\
$\beta_{30}^H$  & 0.14 & 0.06   & 0.04  & 0.11 \\
\hline
$A_L$ & 108.2 & 1.45            & 125.0 & 2.14 \\
$Z_L$ & 42.4  & 1.06            & 48.8  & 1.83 \\
$x^L_{CM}$ [fm] & 0. & 0.14     & 0.    & 0.13 \\
$z^L_{CM}$ [fm] & 10.8 & 0.18  & 10.5  & 0.17 \\
$\beta_{20}^L$  & 0.65 & 0.08   & 0.57  & 0.07 \\
$\beta_{30}^L$  & -0.04 & 0.12  & 0.04  & 0.11 \\
\hline
\hline
\end{tabular}
\caption{Expectation values and standard deviations of several one-body observables 
computed from states representative of the SII and SL fission modes of $^{252}$Cf. 
The first part of the table provides observables of the whole system while the second (resp. third) gives properties of the heavy (resp. light) primary fragment.}
\label{tab:observables1}
\end{table}
%
We find a rather small fluctuation of the elongation of the compound system.
In the asymmetric configuration, the relative standard deviation $\sigma(\beta_{20}) / \langle \beta_{20}\rangle$ is of the order of 2\%.
This is at contrast with smaller systems such as the ground state of $^{20}$Ne for which the elongation shows a relative fluctuation 
$\sigma(\beta_{20}) / \langle \beta_{20}\rangle$ as large as 50\%.
The octupole moment of $^{252}$Cf has a relative fluctuation $\sigma(\beta_{30}) / \langle \beta_{30}\rangle \simeq 12\%$.

\begin{figure}[!h]
    \raggedleft
     \includegraphics[width=1.0\linewidth]{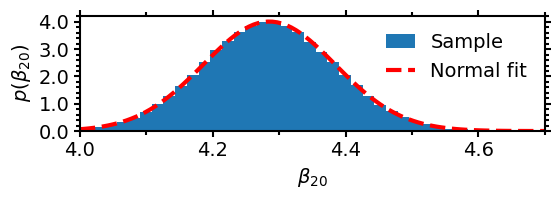}
     \includegraphics[width=1.0\linewidth]{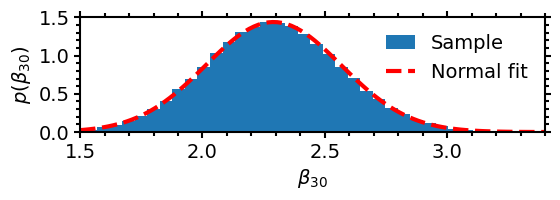}
     \includegraphics[width=1.0\linewidth]{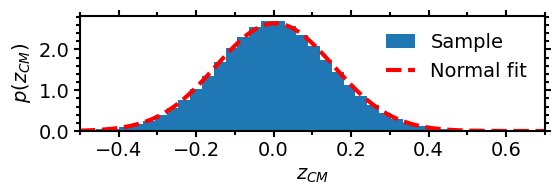}
     \includegraphics[width=0.97\linewidth]{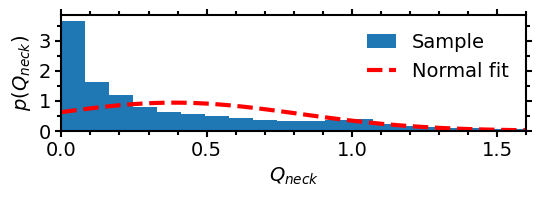}
    \caption{Probability distribution function of several observables of the $^{252}$Cf in a state representative of the SII mode.}
    \label{fig:plot_distrib_01}
\end{figure}

We show in FIG.~\ref{fig:plot_distrib_01} the marginal probability distributions of these deformation variables.
%
%
Our results emphasize a nearly normal behavior for their pdf.
This is likely coming from multipole moments being sums of many random variables nearly independent from each
other, namely the $x^l$, $y^l$ and $z^l$ of the nucleons.
Because HFB calculations break the translational invariance, the position of the center of mass fluctuates
in the particle number projected Bogoliubov vacua under consideration.
Once again due to the large number of particles in the system, the standard deviation
of the coordinates of the center of mass ($x_{CM},y_{CM},z_{CM}$) are only of the order of tenths of fm
and their pdf follows nearly normal laws.
Because we study systems with large elongations along the z-axis, we further find that
the fluctuation of the center of mass is roughly 1.5 times larger along this axis than the 
perpendicular directions.
Finally, we determined the distribution of the $\hat{Q}_{neck}$ observable that characterizes the remaining
number of particles in the neck. 
In both scission configurations, we find a large relative fluctuation ($\simeq 100\%$) along with a non Gaussian 
behavior of the pdf. 
We show in FIG.~\ref{fig:plot_distrib_01} that its is peaked at zero particles with a long
tail that exhausts 90\% of the probability at 1 particle.

\subsection{Fragments deformations}

In the same way as for spins or particle numbers, we extract here the deformations of the
pre-fragments by introducing a Heaviside function $\Theta_{z>z_{neck}}$
into the multipole moment kernels. 
In addition, the origins of the multipole moments are shifted to the expectation values
of the center of mass of each fragment.
The fragments deformations are reported in the second and third parts of TABLE~\ref{tab:observables1}.
As already reported in Refs.~\cite{scamps2018impact}, the SII configuration involves a
slightly octupolar heavy fragment with a significantly elongated light fragment.
This work shows in addition that the shapes of the fragments are prone to 
fluctuations of the order of 0.05 to 0.1 units.

One may attempt to link the shape fluctuation of the fragments to the fluctuation
of the excitation energy of the nascent fragments and ultimately to the standard deviation
of the prompt neutron multiplicity in a given fission mode.
In the regime of mass and deformation of the SII fragments, the constrained HFB calculations reported in the AMEDEE database~\cite{hilaire2007largescale} (obtained with Gogny D1S)
typically show an energy slope of $\Delta E/\Delta \beta_{20} \simeq 25$ MeV.
Using this rough estimate of the deformation energy, the shape fluctuation of the fragments
would imply a 1-2 MeV standard deviation of the fragments excitation energy.
A similar order of magnitude is found in the SL mode with $\Delta E/\Delta \beta_{20} \simeq 18 $MeV.
Such a fluctuation in the excitation energy would only explain a fluctuation of $\simeq$ 0.2 prompt neutrons per fragment.
For comparison, the typical standard deviation of the total prompt neutron multiplicity is of the order of 0.9 neutron for SII and close to 1.4 for SL~\cite{kalinin2002measurements,dushin2004facility}.
As a conclusion, the fluctuation in the shape of fragments only explains a small fraction of 
the fluctuation in the excitation energy.
Other contributors to the excitation energy related to intrinsic degrees of freedom may explain the missing fluctuation.
Such a difference may also come from the total energy of the state under consideration that is smaller that the ground state energy of the californium,
or the fact that a mean-field wavefunction simply underestimates the shape fluctuations.

\subsection{Inter-fragments interaction energies}

In the wavefunctions considered, the fragments still interact through
the Coulomb repulsion and a residual nuclear interaction.
The expectation values of the corresponding two-body operators have already been discussed in~\cite{younes2009microscopic,simenel2014formation,carpentier2024microscopic}.
In this section, we estimate the order of magnitude for the fluctuation of the interaction energy between
the fragments and compare it to the experimentally known fluctuation in the total kinetic energy.

With the same philosophy as for the fragments one-body observables, we can express the inter-fragment interaction energy of any two-body interaction diagonal in the 
space/intrinsic spin representation.
\begin{equation}
    \hat{V}^{inter} = 
    \frac{1}{2}\int_{rr'} \left[2 \, v(r,r') \Theta_{z<z_{neck}} \Theta_{z'>z_{neck}}\right] 
    c^\dagger_rc^\dagger_{r'} c_{r'} c_r.
    \label{eq:inter}
\end{equation}
We introduced here two Heaviside functions to only account for contributions involving one particle of the left fragment with one particle of the right fragment.
Note also that the matrix element of the interaction $v(r,r')$ is multiplied by two to take into account the potential energy felt by the left fragment and the potential energy
felt by the right fragment.
In what follows, we apply this formula with the Coulomb interaction
\begin{equation}
    v_{coul}(r,r') = 
    \frac{e^2}{4\pi\epsilon_0}\frac{1}{|r-r'|} 
    \label{eq:coulomb}
\end{equation}
to estimate the inter-fragment Coulomb repulsion close to scission.
For the nuclear part, it is not possible to directly use the Gogny D1M effective interaction due to the non-diagonal
terms that it contains.
In the current work, we circumvent this difficulty by relying on the simple spin and isospin independent diagonal Yukawa nuclear potential
\begin{equation}
    v_{nucl}(r,r') = -V_0 \frac{e^{-|r-r'|/r_0}}{|r-r'|/r_0}.
    \label{eq:yukawa}
\end{equation}
Following \cite{ring2004nuclear} we parameterize this central force with $V_0=-50$ MeV and $r_0=1.4$ fm.
These constants yield an expectation value of -15.4 MeV for the residual nuclear interaction in our SII state.
This seems consistent with the -20 MeV and -26 MeV obtained for $^{240}$Pu in Refs.~\cite{younes2011nuclear} and \cite{carpentier2024microscopic}, respectively.
Note that a strict comparison to the aforementioned studies remains delicate as the 
fragment separation method used is different from the spatial treatment used here in Eq.\eqref{eq:inter}.
The definitions \eqref{eq:inter},\eqref{eq:coulomb},\eqref{eq:yukawa} along with the kernel~\eqref{eq:2bdykernel} allows us to estimate the pdf of the fragments interaction energies.

We report in TABLE~\ref{tab:energies} the expectation values and standard deviations for the Coulomb, nuclear and total interaction
potentials between the fragments.
%
\begin{table}[h!]
\begin{tabular}{l|ll|ll}
\hline
\hline
 & SII & & SL & \\
 & $\langle\hat{O}\rangle$ & $\sigma(\hat{O})$ & $\langle\hat{O}\rangle$ & $\sigma(\hat{O})$ \\
  \hline
$V^{inter}_{coul}$  [MeV] & 184.1 & 4.1    & 172.7 & 3.6 \\
$V^{inter}_{nuc}$   [MeV] & -15.4 & 7.2    & -8.6  & 5.4 \\
$V^{inter}_{tot}$   [MeV] & 168.7 & 7.2    & 164.1 & 5.8 \\
\hline
TKE [MeV] \cite{gook2014prompt}  & 184.7 & 8.8 & 186.0 & 11.6 \\
\hline
\hline
\end{tabular}
\caption{Expectation values and standard deviations of the fragments interaction energies 
in states representative of the SII and SL fission modes of $^{252}$Cf.
We compare it to the experimental total kinetic energy (TKE) taken from \cite{gook2014prompt}, 
considering a light fragment of mass 108 for the SII mode and 126 for SL mode.}
\label{tab:energies}
\end{table}
%
Close to scission, we expect the total energy of the system to be shared between the Coulomb repulsion, 
a residual nuclear interaction between the fragments, an excitation energies of the fragments ($\text{TXE}_{s}$) and a kinetic energy ($\text{TKE}_s$) .
After full acceleration of the fragments but before the prompt particle emission, the energy should re-organize into
the measured total kinetic energy (TKE) along with a total excitation energy (TXE) of the fragments.
\begin{equation}
V^{inter}_{tot}  + \text{TKE}_{s} + \text{TXE}_{s} = \text{TKE} + \text{TXE}
\end{equation}
Making the bold assumption that the total excitation energy does not change much from the scission configuration to the fully accelerated one,
our results yield $\text{TKE}_s \simeq 16$ MeV of kinetic energy in the scission configuration.
This seems compatible with time-dependent mean-field predictions of Ref.~\cite{simenel2014formation} giving us confidence 
in the order of magnitude of the different terms at stake.

\begin{figure}[!h]
    \raggedleft
     \includegraphics[width=1.0\linewidth]{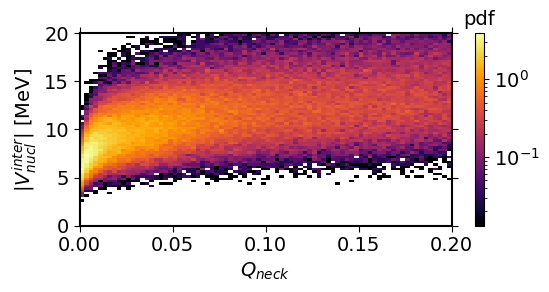}
     \caption{Bi-dimensional pdf of the number of particles in the neck (estimated with $Q_{neck}$) and the magnitude of the nuclear inter-fragment interaction $V_{^{inter}_{nucl}}$ for the SII scission configuration of $^{252}$Cf.}
    \label{fig:plot_qneck_vinter}
\end{figure}

We now look at the fluctuations of the different energy reservoirs.
Our results suggest that a large part of the measured fluctuation of the kinetic energy originates from the fluctuation 
of the interaction potential between the fragments close to scission.
In our model, the later originates from the fluctuation in the positions of the nucleons.
This is especially valid for the SII mode for which the standard deviation of the residual interaction is within 1 MeV
of the standard deviation of the total kinetic energy.
In addition, we emphasize that the main contributor to this fluctuation is the nuclear interaction and not the Coulomb repulsion.
The small range of the nuclear interaction suggests that such variability of the inter-fragment nuclear interaction
should be connected to the position of the few nucleons close to the neck.
We show in FIG.~\ref{fig:plot_qneck_vinter} the bi-dimensional pdf of $Q_{neck}$ and the inter-fragment nuclear interaction $V_{^{inter}_{nucl}}$.
%
%
We see indeed that the two observables are strongly inter-dependent. 
Quantitatively, the magnitude of the nuclear interaction presents a correlation coefficient of 0.75 with the neck operator value.
In the other hand, the inter-fragment Coulomb interaction is less sensitive to the position of nucleons of the neck with a correlation coefficient of 0.1 with $Q_{neck}$.

\begin{figure}[!h]
    \raggedleft
     \includegraphics[width=1.0\linewidth]{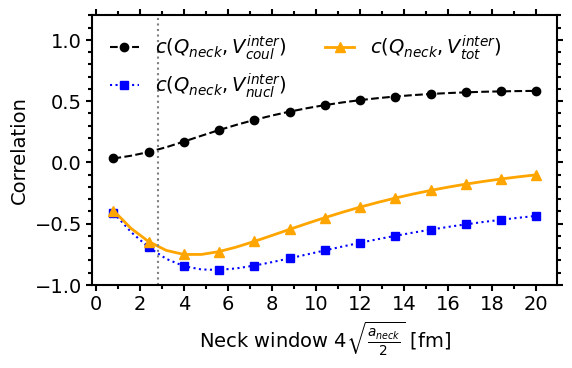}
     \caption{Correlation between the neck operator value and the inter-fragment interactions as a function of the spatial window used to estimate $Q_{neck}$. We show the inter-fragment Coulomb, nuclear and total interactions in the SII scission configuration of a $^{252}$Cf. The grey vertical line corresponds to our reference calculation with $a_{neck}=1$.}
    \label{fig:plot_correlations}
\end{figure}

In this analysis, both the fragments properties and the value of $Q_{neck}$ depend on the rather arbitrary splitting position $z_{neck}$. In a real fission event, the dependency to this choice becomes negligible as the fragments move away from each other. In the current calculation varying $z_neck$ by 1 fm on the right or left typically modifies the nuclear and Coulomb inter-fragments interactions by 2 MeV. The anti-correlation between $Q_{neck}$ and the total fragment interaction is robust to this change and remains above 0.6.
With a similar philosophy we checked the sensitivity of our results to the size of the neck that we consider. This size is encoded by the parameter $a_{neck}$ involved in the definition of the neck operator. We plot in Fig.~\ref{fig:plot_correlations} the correlations between the neck operator and the inter-fragments interactions for a range $a_{neck}$. The x-axis actually represents the spatial window exhausting 95\% of the Gaussian form factor involved in $Q_{neck}$, namely $4\sqrt{ a_{neck}/2 }$.
%
%
Our major conclusion about a strong anti-correlation between the presence of particles in the neck region and the nuclear interaction between the fragments holds.
This picture emphasizes also a minimum in the correlation $c(Q_{neck}, V^{inter}_{nucl})$ as a function of the neck window considered.
This indicates that particles localized in the tips of the nascent fragments
play the largest role in the inter-fragment interaction.
Because pairing is a major driver of the fluctuation of matter at the nuclear surface,
it may be strongly correlated to the fluctuations measured in the kinetic energy
of the fragments.

To put it in a nutshell, this study shows that within our EDF model, most of the fluctuation of the total kinetic energy of the fragments
takes its origin in the fluctuation of the nuclear inter-fragment interaction close to scission.
In the space/intrinsic spin representation, the later is mostly induced by the fluctuation of the positions of a few nucleons in the neck region.

\subsection{Torques acting on the fragments}

The mechanisms at the origin of the fission fragments angular momentum has been in the spotlight during the last years~\cite{bulgac2022angular,scamps2023spatial,scamps2024quantal,francheteau2024scission}.
It may seem counter-intuitive that mean-field calculations with axial one-body densities lead to non-zero spin components of the fragments
perpendicular to the fission axis.
As mentioned in the introduction, the quantum fluctuation of the nucleons positions provides a 
pedagogical interpretation of this phenomena.
In Ref.~\cite{scamps2025uncertainty}, the authors connect the angular momentum content of the fragments close to scission
to the fluctuation in the orientation of the principal deformation axis of the fragments.
In this section, we focus on the residual nuclear and Coulomb torques acting on the fragments in states close to scission.
During the dynamics from scission configuration to fully accelerated fragments, we expect that such a residual torque 
will provide an additional angular momentum contribution to the fragments.

The torque felt by a fragment and resulting from its interaction with the other fragment 
can be encoded as a vector 2-body observables of the form~\eqref{eq:inter}.
The vector of kernels involved takes the form
\begin{equation}
    \bm{\tau}(r,r') = (\bm{r}-\bm{r_0})\times \bm{f(\bm{r},\bm{r}')}
\end{equation}
with
\begin{align}
\bm{f}_{coul}(r_i, r_j) &= \frac{e^2}{4\pi\epsilon_0}\frac{1}{(r_{i}-r_j)^2} \bm{e}_{ij}, \\
\bm{f}_{nucl}(r_i, r_j) &= -V_0 \, \text{exp}\left(-\frac{d}{r_0}\right) \frac{(d+r_0)}{d^2} \bm{e}_{ij}.
\end{align}
Here $\bm{e}_{ij}$ denotes the direction of the force felt by the nucleon $j$ from the nucleon $i$ and $\bm{r_0}$ is the center of mass of the fragment considered.
The amplitudes of the forces correspond to the spatial gradient of their respective potentials~\eqref{eq:coulomb} and \eqref{eq:yukawa}.

\begin{figure}[!h]
    \raggedleft
    \includegraphics[width=1.0\linewidth]{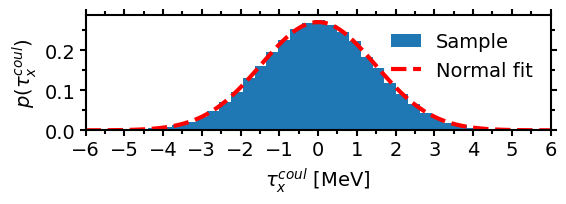}
     \includegraphics[width=1.0\linewidth]{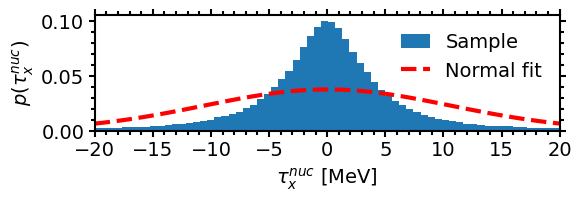}
      \caption{Pdf of the x-component of the Coulomb (top) and nuclear (bottom) torques induced by the heavy fragment on the light fragment the state representative of the SII fission mode of $^{252}$Cf.}
    \label{fig:plot_distrib_torques}
\end{figure}

Based on our sample of configurations, we compute the pdfs of the
torque components for the Coulomb, nuclear and total interactions.
We summarize the expectation values and standard deviations obtained in TABLE~\ref{tab:torques}.
%
\begin{table}[h!]
\begin{tabular}{l|ll|ll}
\hline
\hline
 & SII & & SL & \\
 & $\langle\hat{O}\rangle$ & $\sigma(\hat{O})$ & $\langle\hat{O}\rangle$ & $\sigma(\hat{O})$ \\
  \hline
$\bm{\tau}_x^{coul, H}$  & 0.   & 1.3      & 0 & 1.2 \\
$\bm{\tau}_x^{nucl, H}$   & 0.   & 10.4     & 0 & 12.7 \\
$\bm{\tau}_x^{tot, H}$     & 0.   & 8.7    & 0 & 10.2 \\
\hline
$\bm{\tau}_x^{coul, L}$  & 0.   & 1.5      & 0 & 1.2 \\
$\bm{\tau}_x^{nucl, L}$   & 0.   & 10.6     & 0  & 14.2 \\
$\bm{\tau}_x^{tot, L}$     & 0.   & 8.7    & 0 & 5.8 \\
\hline
\hline
\end{tabular}
\caption{Expectation values and standard deviations of the x-axis component of the residual torques felt by the heavy (H) and light (L) fragments 
in states representative of the SII and SL scission of $^{252}$Cf.
We provide the torque coming from the Coulomb interaction, the nuclear interaction and the sum of the two.
Values are provided in $\hbar.\text{zs}^{-1}$ units.}
\label{tab:torques}
\end{table}
%
The expectation values of the torque in every directions are zero up to the statistical uncertainty.
On the other hand, we see significant standard deviation for the torque perpendicular to 
the z-axis induced by the fluctuation of the nucleons positions.
In a similar way as for the interaction energies, most of the fluctuation of the torque takes its origin
in the nuclear interaction as opposed to the Coulomb repulsion.
This leads us to the conclusion that the residual torque is mostly induced by the few particles 
close to the neck area.
The figure~\ref{fig:plot_distrib_torques} further illustrates the results obtained for the torque component perpendicular to the z-axis
and induced by the heavy fragment on the light fragment in the state representative of the SII mode.
%
%
On the one hand, the Coulomb torque pdf follows a normal behavior resulting from the sum over the numerous pairs of protons that contribute.
On the other hand, the residual nuclear torque pdf is not Gaussian at all.
Overall, the residual torque is mostly oriented perpendicular to the fission axis. The z-axis component represents only a few percents (resp. 10 to 20 percents) of the x or y axis
for the Coulomb part (resp. nuclear part).

In a classical picture, the magnitude of the torques integrated over $\simeq$1 zeptosecond would give an estimate of the additional angular momentum
induced by the residual inter-fragment torque.
The zeptosecond corresponds to the typical time for the fragments to further separate leading to a vanishing residual nuclear interaction.
This integration yields an additional fluctuation of 8$\hbar$ in the angular 
momentum to be added to the 
one present at scission.
Such a naive picture seems to overestimate the width of the final angular momentum of the primary fragments owing to the width of the angular momentum 
that is already present in the scission configurations (\textit{i.e.} 6-12 $\hbar$ from \cite{scamps2025uncertainty}).
This mismatch could be attributed to several causes such as the classical equation of motion for the angular momentum assumed here
or an overestimation of the angular momentum width at scission from mean-field wavefunctions.

To conclude, this work predicts a significant residual torque between the fragments close to scission.
It is premature to assess wether this residual torque actually plays a significant role in the spin generation mechanism
as compared to the spin already present close to scission.
In any case, we show that such residual torque is mostly coming from the nuclear interaction acting between the few nucleons in the neck region.

\section{Conclusion}
\label{conclusion}

In this study we propose a method to sample the nucleons positions and intrinsic spin from the many-body density 
of a Bogoliubov vacuum projected onto good particle number.
This algorithm provides a pedagogical visualization of the fluctuating cloud of nucleons.
In addition, it enables the computation of the pdf of many-body observables that are diagonal
in the position/intrinsic spin representation as well as their correlations.
We demonstrated that such a sampling is doable and converges even for heavy nuclei.

In the context of fission, it allowed us to estimate the pdf of many obervables related to the geometry 
of the compound nucleus close to its scission.
Multipole moments of the compound system and of the fragments show normal behavior related to the 
large number of particles of the system. 
This is at contrast with the number of particles in the neck which has an expectation value below 
one particle with a large fluctuation.
This study also enabled prediction of two-body observables pdfs giving the opportunity to look into the flucutation of the 
inter-fragment interaction.
In particular, we found that a large part of the measured total kinetic energy fluctuation of the fragment
likely takes its origin in the fluctuation of the residual nuclear interaction close to scission.
Similar reasoning shows a residual torque between the fragments that is induced by the nuclear interaction between 
the few nucleons present in the neck region.

To better ground these conclusions, a systematic study of the nuclear interaction fluctuation 
on a variety of wavefunctions close to scission should be carried on.
This would pave the way to understand how the fluctuation varies with the disappearance of the 
neck and how our results change for a compact fission mode.
A similar work could also be performed starting from a snapshot of a time-dependent EDF simulation.
Comparing to the present results, such a study would shade light on the impact of excitation energy
and pre-scission kinetic energy on the fluctuation of several observables.
Performing systematic studies on heavy systems would
probably require the usage of more advanced Markov chain algorithms such as the DE-Metropolis~\cite{braak2006markov}.
Finally, we could attempt to generalize this approach to non-diagonal observables
in the position/intrinsic spin representation following what is done in the quantum Monte Carlo community~\cite{lynn2019quantum}. 
This would typically enable one to obtain directly the pdfs of the kinetic energies
and of a more realistic nuclear interaction.

\begin{acknowledgements}
The authors would like to thank
G.~Belier, A. Francheteau, D.~Lacroix, G. Scamps, and M. Verriere for stimulating discussions.
\end{acknowledgements}

\bibliographystyle{apsrev4-1}
\bibliography{biblio.bib}

@article{baye2015lagrangemesh,
  title = {The {{Lagrange-mesh}} Method},
  author = {Baye, Daniel},
  year = 2015,
  journal = {Physics Reports},
  series = {The {{Lagrange-mesh}} Method},
  volume = {565},
  pages = {1--107},
  doi = {10.1016/j.physrep.2014.11.006}
}

@article{DIETRICH2010249,
title = {Microscopic transport theory of nuclear processes},
journal = {Nuclear Physics A},
volume = {832},
number = {3},
pages = {249-288},
year = {2010},
issn = {0375-9474},
doi = {https://doi.org/10.1016/j.nuclphysa.2009.11.004},
url = {https://www.sciencedirect.com/science/article/pii/S0375947409008951},
author = {K. Dietrich and J.-J. Niez and J.-F. Berger}
}

@article{bernard2011microscopic,
  title = {Microscopic and Nonadiabatic {{Schr\"odinger}} Equation Derived from the Generator Coordinate Method Based on Zero- and Two-Quasiparticle States},
  author = {Bernard, R. and Goutte, H. and Gogny, D. and Younes, W.},
  year = 2011,
  journal = {Phys. Rev. C},
  volume = {84},
  number = {4},
  pages = {044308},
  doi = {10.1103/PhysRevC.84.044308}
}

@misc{bjelcic2025excitation,
  title = {Excitation Energy of Fission Fragments within Nuclear Time-Dependent Density Functional Theory},
  author = {Bjel{\v c}i{\'c}, Antonio and Schunck, Nicolas and Verriere, Marc},
  year = 2025,
  number = {arXiv:2510.06701},
  primaryclass = {nucl-th},
  publisher = {arXiv},
  doi = {10.48550/arXiv.2510.06701},
  archiveprefix = {arXiv}
}

@article{bohr1939mechanism,
  title = {The {{Mechanism}} of {{Nuclear Fission}}},
  author = {Bohr, N. and Wheeler, J. A.},
  year = 1939,
  journal = {Phys. Rev.},
  volume = {56},
  number = {5},
  pages = {426--450},
  doi = {10.1103/PhysRev.56.426}
}

@article{braak2006markov,
  title = {A {{Markov Chain Monte Carlo}} Version of the Genetic Algorithm {{Differential Evolution}}: Easy {{Bayesian}} Computing for Real Parameter Spaces},
  author = {Braak, Cajo J.},
  year = 2006,
  journal = {Statistics and Computing},
  volume = {16},
  number = {3},
  pages = {239--249},
  doi = {10.1007/s11222-006-8769-1}
}

@article{brooks1998general,
  title = {General {{Methods}} for {{Monitoring Convergence}} of {{Iterative Simulations}}},
  author = {Brooks, Stephen P. and Gelman, Andrew},
  year = 1998,
  journal = {Journal of Computational and Graphical Statistics},
  volume = {7},
  number = {4},
  pages = {434--455},
  publisher = {Taylor \& Francis},
  doi = {10.1080/10618600.1998.10474787}
}

@article{brosa1990nuclear,
  title = {Nuclear Scission},
  author = {Brosa, Ulrich and Grossmann, Siegfried and M{\"u}ller, Andreas},
  year = 1990,
  journal = {Physics Reports},
  volume = {197},
  number = {4},
  pages = {167--262},
  doi = {10.1016/0370-1573(90)90114-H}
}

@article{bulgac2016induced,
  title = {Induced {{Fission}} of {\textsuperscript{240}}{{Pu}} within a {{Real-Time Microscopic Framework}}},
  author = {Bulgac, Aurel and Magierski, Piotr and Roche, Kenneth J. and Stetcu, Ionel},
  year = 2016,
  journal = {Phys. Rev. Lett.},
  volume = {116},
  number = {12},
  pages = {122504},
  doi = {10.1103/PhysRevLett.116.122504}
}

@article{bulgac2019fission,
  title = {Fission Dynamics of {{240Pu}} from Saddle to Scission and Beyond},
  author = {Bulgac, Aurel and Jin, Shi and Roche, Kenneth J. and Schunck, Nicolas and Stetcu, Ionel},
  year = 2019,
  journal = {Phys. Rev. C},
  volume = {100},
  number = {3},
  pages = {034615},
  publisher = {American Physical Society},
  doi = {10.1103/PhysRevC.100.034615}
}

@article{bulgac2020fissionfragment,
  title = {Fission-Fragment Excitation Energy Sharing beyond Scission},
  author = {Bulgac, Aurel},
  year = 2020,
  journal = {Phys. Rev. C},
  volume = {102},
  number = {4},
  pages = {044609},
  publisher = {American Physical Society},
  doi = {10.1103/PhysRevC.102.044609}
}

@article{bulgac2022angular,
  title = {Angular Correlation between the Fission Fragment Intrinsic Spins},
  author = {Bulgac, Aurel},
  year = 2022,
  journal = {Phys. Rev. C},
  volume = {106},
  number = {1},
  pages = {014624},
  publisher = {American Physical Society},
  doi = {10.1103/PhysRevC.106.014624}
}

@article{bulgac2025timedependent,
  title = {Time-{{Dependent Density Functional Theory Description}} of \$\textasciicircum\textbraceleft 238\textbraceright\textbackslash mathrm\textbraceleft{{U}}\textbraceright (\textbackslash mathrm\textbraceleft n\textbraceright,\textbackslash mathrm\textbraceleft f\textbraceright )\$, \$\textasciicircum\textbraceleft 240,242\textbraceright\textbackslash mathrm\textbraceleft{{Pu}}\textbraceright (\textbackslash mathrm\textbraceleft n\textbraceright,\textbackslash mathrm\textbraceleft f\textbraceright )\$, and \$\textasciicircum\textbraceleft 237\textbraceright\textbackslash mathrm\textbraceleft{{Np}}\textbraceright (\textbackslash mathrm\textbraceleft n\textbraceright,\textbackslash mathrm\textbraceleft f\textbraceright )\$ {{Reactions}}},
  author = {Bulgac, Aurel and Abdurrahman, Ibrahim and Kafker, Matthew and Stetcu, Ionel},
  year = 2025,
  journal = {Phys. Rev. Lett.},
  volume = {135},
  number = {6},
  pages = {062501},
  publisher = {American Physical Society},
  doi = {10.1103/2k8k-vpng}
}

@article{carpentier2024construction,
  title = {Construction of {{Continuous Collective Energy Landscapes}} for {{Large Amplitude Nuclear Many-Body Problems}}},
  author = {Carpentier, Paul and Pillet, Nathalie and Lacroix, Denis and Dubray, No{\"e}l and Regnier, David},
  year = 2024,
  journal = {Phys. Rev. Lett.},
  volume = {133},
  number = {15},
  pages = {152501},
  publisher = {American Physical Society},
  doi = {10.1103/PhysRevLett.133.152501}
}

@phdthesis{carpentier2024microscopic,
  title = {Microscopic and Dynamical Description of the Fission Process Including Intrinsic Excitations},
  author = {Carpentier, Paul},
  year = 2024,
  langid = {english},
  school = {Universit\'e Paris-Saclay},
  url = {https://theses.hal.science/tel-04761482v1}
}

@misc{dubray2025hfb3,
  title = {{{HFB3}}: An Axial {{HFB}} Solver with {{Gogny}} Forces Using a 2-Center {{HO}} Basis ({{C}}++/{{Python}})},
  author = {Dubray, N. and Ebran, J. P. and Carpentier, P. and Frosini, M. and Zdeb, A. and Pillet, N. and Newsome, J. and Verri{\`e}re, M. and Accorto, G. and Regnier, D.},
  year = 2025,
  number = {arXiv:2506.10745},
  primaryclass = {nucl-th},
  publisher = {arXiv},
  doi = {10.48550/arXiv.2506.10745},
  archiveprefix = {arXiv}
}

@article{dushin2004facility,
  title = {Facility for Neutron Multiplicity Measurements in Fission},
  author = {Dushin, V.N. and Hambsch, F.-J. and Jakovlev, V.A. and Kalinin, V.A. and Kraev, I.S. and Laptev, A.B. and Nikolaev, D.V. and Petrov, B.F. and Petrov, G.A. and Petrova, V.I. and Pleva, Y.S. and Shcherbakov, O.A. and Shpakov, V.I. and Sokolov, V.E. and Vorobyev, A.S. and Zavarukhina, T.A.},
  year = 2004,
  journal = {Nuclear Instruments and Methods in Physics Research Section A: Accelerators, Spectrometers, Detectors and Associated Equipment},
  volume = {516},
  number = {2--3},
  pages = {539--553},
  doi = {10.1016/j.nima.2003.09.029}
}

@book{fishman1996monte,
  title = {Monte {{Carlo}}},
  author = {Fishman, G.},
  year = 1996,
  edition = {Corrected},
  publisher = {Springer},
  isbn = {0-387-94527-X}
}

@article{fomenko1970structure,
  title = {On the Structure of the Paired States},
  author = {Fomenko, V. N.},
  year = 1970,
  journal = {J. Phys. A: Gen. Phys.},
  volume = {3},
  number = {5},
  pages = {520},
  doi = {10.1088/0305-4470/3/5/008},
  langid = {english}
}

@article{francheteau2024scission,
  title = {Scission {{Deformation}} of the \$\textasciicircum\textbraceleft 120\textbraceright\textbackslash mathrm\textbraceleft{{Cd}}\textbraceright/\textasciicircum\textbraceleft 132\textbraceright\textbackslash mathrm\textbraceleft{{Sn}}\textbraceright\$ {{Neutronless Fragmentation}} in \$\textasciicircum\textbraceleft 252\textbraceright\textbackslash mathrm\textbraceleft{{Cf}}\textbraceright (\textbackslash mathrm\textbraceleft sf\textbraceright )\$},
  author = {Francheteau, A. and Gaudefroy, L. and Scamps, G. and Roig, O. and M{\'e}ot, V. and Ebran, A. and B{\'e}lier, G.},
  year = 2024,
  journal = {Phys. Rev. Lett.},
  volume = {132},
  number = {14},
  pages = {142501},
  publisher = {American Physical Society},
  doi = {10.1103/PhysRevLett.132.142501}
}

@article{gelman1992inference,
  title = {Inference from {{Iterative Simulation Using Multiple Sequences}}},
  author = {Gelman, Andrew and Rubin, Donald B.},
  year = 1992,
  journal = {Statistical Science},
  volume = {7},
  number = {4},
  pages = {457--472},
  publisher = {Institute of Mathematical Statistics},
  doi = {10.1214/ss/1177011136}
}

@article{gook2014prompt,
  title = {Prompt Neutron Multiplicity in Correlation with Fragments from Spontaneous Fission of \$\textasciicircum\textbraceleft 252\textbraceright\textbackslash mathrm\textbraceleft{{Cf}}\textbraceright\$},
  author = {G{\"o}{\"o}k, A. and Hambsch, F.-J. and Vidali, M.},
  year = 2014,
  journal = {Phys. Rev. C},
  volume = {90},
  number = {6},
  pages = {064611},
  publisher = {American Physical Society},
  doi = {10.1103/PhysRevC.90.064611}
}

@article{goriely2009first,
  title = {First {{Gogny-Hartree-Fock-Bogoliubov Nuclear Mass Model}}},
  author = {Goriely, S. and Hilaire, S. and Girod, M. and P{\'e}ru, S.},
  year = 2009,
  journal = {Phys. Rev. Lett.},
  volume = {102},
  number = {24},
  pages = {242501},
  doi = {10.1103/PhysRevLett.102.242501},
  langid = {english}
}

@article{hilaire2007largescale,
  title = {Large-Scale Mean-Field Calculations from Proton to Neutron Drip Lines Using the {{D1S Gogny}} Force},
  author = {Hilaire, S. and Girod, M.},
  year = 2007,
  journal = {Eur. Phys. J. A},
  volume = {33},
  number = {2},
  pages = {237--241},
  doi = {10.1140/epja/i2007-10450-2},
  copyright = {\copyright{} Societ\`a Italiana di Fisica and Springer-Verlag, 2007},
  langid = {english}
}

@misc{introduction,
  title = {Introduction to Mathematical Statistics \textbar{} {{Robert V}}. {{Hogg}}, {{Allen T}}. {{Craig}} \textbar{} Digital Library Bookzz},
  howpublished = {http://bookzz.org/book/448462/a59fe3}
}

@article{kalinin2002measurements,
  title = {Measurements of {{Prompt Neutron Multiplicity Distributions}} in {{Correlation}} with {{Mass-Energy Distribution}} of {{Fission Fragments}} in {{Spontaneous Fission}} of {{252Cf}}, {{244Cm}} and {{248Cm}}},
  author = {Kalinin, Valeri and Dushin, Victor and Hambsch, Franz-Josef and Jakovlev, Vladimir and Kraev, Il'ya and Laptev, Alexander and Petrov, Boris and Petrov, Guennady and Pleva, Yuri and Shcherbakov, Oleg and Sokolov, Vladimir and Vorobyev, Alexander},
  year = 2002,
  journal = {Journal of Nuclear Science and Technology},
  volume = {39},
  number = {sup2},
  pages = {250--253},
  publisher = {Taylor \& Francis},
  doi = {10.1080/00223131.2002.10875086}
}

@article{koonin1997shell,
  title = {Shell Model {{Monte Carlo}} Methods},
  author = {Koonin, S. E. and Dean, D. J. and Langanke, K.},
  year = 1997,
  journal = {Physics Reports},
  volume = {278},
  number = {1},
  pages = {1--77},
  doi = {10.1016/S0370-1573(96)00017-8}
}

@article{lasseri2024generative,
  title = {Generative Deep-Learning Reveals Collective Variables of {{Fermionic}} Systems},
  author = {Lasseri, Rapha{\"e}l-David and Regnier, David and Frosini, Mika{\"e}l and Verriere, Marc and Schunck, Nicolas},
  year = 2024,
  journal = {Phys. Rev. C},
  volume = {109},
  number = {6},
  pages = {064612},
  publisher = {American Physical Society},
  doi = {10.1103/PhysRevC.109.064612}
}

@article{lau2022smoothing,
  title = {Smoothing of One- and Two-Dimensional Discontinuities in Potential Energy Surfaces},
  author = {Lau, N.-W. T. and Bernard, R. N. and Simenel, C.},
  year = 2022,
  journal = {Phys. Rev. C},
  volume = {105},
  number = {3},
  pages = {034617},
  publisher = {American Physical Society},
  doi = {10.1103/PhysRevC.105.034617}
}

@article{li2023generalizedb,
  title = {Generalized Time-Dependent Generator Coordinate Method for Small- and Large-Amplitude Collective Motion},
  author = {Li, B. and Vretenar, D. and Nik{\v s}i{\'c}, T. and Zhao, P. W. and Meng, J.},
  year = 2023,
  journal = {Phys. Rev. C},
  volume = {108},
  number = {1},
  pages = {014321},
  publisher = {American Physical Society},
  doi = {10.1103/PhysRevC.108.014321}
}

@article{li2024generalized,
  title = {Generalized Time-Dependent Generator Coordinate Method for Induced Fission Dynamics},
  author = {Li, B. and Vretenar, D. and Nik{\v s}i{\'c}, T. and Zhao, J. and Zhao, P. W. and Meng, J.},
  year = 2024,
  journal = {Front. Phys.},
  volume = {19},
  number = {4},
  primaryclass = {nucl-th},
  pages = {44201},
  doi = {10.1007/s11467-023-1381-4},
  archiveprefix = {arXiv}
}

@article{li2025fourier,
  title = {Fourier Shape Parameterization in Covariant Density Functional Theory for Nuclear Fission},
  author = {Li, Zeyu and Su, Yang and Liu, Lile and Chen, Yongjing and Li, Zhipan},
  year = 2025,
  journal = {Physics Letters B},
  volume = {866},
  pages = {139509},
  doi = {10.1016/j.physletb.2025.139509}
}

@article{li2025microscopica,
  title = {Microscopic Model for Yields and Total Kinetic Energy in Nuclear Fission},
  author = {Li, B. and Vretenar, D. and Nik{\v s}i{\'c}, T. and Zhao, P. W. and Meng, J.},
  year = 2025,
  journal = {Phys. Rev. C},
  volume = {111},
  number = {5},
  pages = {L051302},
  publisher = {American Physical Society},
  doi = {10.1103/PhysRevC.111.L051302}
}

@article{lynn2019quantum,
  title = {Quantum {{Monte Carlo Methods}} in {{Nuclear Physics}}: {{Recent Advances}}},
  author = {Lynn, J. E. and Tews, I. and Gandolfi, S. and Lovato, A.},
  year = 2019,
  journal = {Annual Review of Nuclear and Particle Science},
  volume = {69},
  number = {Volume 69, 2019},
  pages = {279--305},
  publisher = {Annual Reviews},
  doi = {10.1146/annurev-nucl-101918-023600},
  langid = {english}
}

@article{marevic2020fission,
  ids = {marevic2020fissiona},
  title = {Fission of \$\textasciicircum\textbraceleft 240\textbraceright\textbackslash mathrm\textbraceleft{{Pu}}\textbraceright\$ with {{Symmetry-Restored Density Functional Theory}}},
  author = {Marevi{\'c}, P. and Schunck, N.},
  year = 2020,
  journal = {Phys. Rev. Lett.},
  volume = {125},
  number = {10},
  pages = {102504},
  publisher = {American Physical Society},
  doi = {10.1103/PhysRevLett.125.102504}
}

@article{marevic2023quantum,
  title = {Quantum Fluctuations Induce Collective Multiphonons in Finite {{Fermi}} Liquids},
  author = {Marevi{\'c}, Petar and Regnier, David and Lacroix, Denis},
  year = 2023,
  journal = {Phys. Rev. C},
  volume = {108},
  number = {1},
  pages = {014620},
  publisher = {American Physical Society},
  doi = {10.1103/PhysRevC.108.014620}
}

@misc{marevic2025microscopic,
  title = {Microscopic Theory of Angular Momentum Distributions across the Full Range of Fission Fragments},
  author = {Marevi{\'c}, Petar and Schunck, Nicolas and Verriere, Marc},
  year = 2025,
  number = {arXiv:2506.10777},
  primaryclass = {nucl-th},
  publisher = {arXiv},
  doi = {10.48550/arXiv.2506.10777},
  archiveprefix = {arXiv}
}

@article{matsumoto2022visualization,
  title = {Visualization of Nuclear Many-Body Correlations with the Most Probable Configuration of Nucleons},
  author = {Matsumoto, Moemi and Tanimura, Yusuke},
  year = 2022,
  journal = {Phys. Rev. C},
  volume = {106},
  number = {1},
  pages = {014307},
  publisher = {American Physical Society},
  doi = {10.1103/PhysRevC.106.014307}
}

@article{meitner1939disintegration,
  title = {Disintegration of {{Uranium}} by {{Neutrons}}: A {{New Type}} of {{Nuclear Reaction}}},
  author = {Meitner, L. and Frisch, O.R},
  year = 1939,
  journal = {Nat.},
  volume = {143},
  pages = {239--240}
}

@article{morfouace2025asymmetric,
  title = {An Asymmetric Fission Island Driven by Shell Effects in Light Fragments},
  author = {Morfouace, P. and Taieb, J. and Chatillon, A. and Audouin, L. and Blanchon, G. and Bernard, R. N. and Dubray, N. and Pillet, N. and Regnier, D. and {Alvarez-Pol}, H. and Amjad, F. and Andr{\'e}, P. and Authelet, G. and Atar, L. and Aumann, T. and Benlliure, J. and Boretzky, K. and Bott, L. and Brecelj, T. and Caesar, C. and Carpentier, P. and Casarejos, E. and Cederk{\"a}ll, J. and Corsi, A. and {Cortina-Gil}, D. and Cvetinovi{\'c}, A. and Filippo, E. De and Dickel, T. and Feijoo, M. and Fonseca, L. M. and Galaviz, D. and {Garc{\'i}a-Jim{\'e}nez}, G. and Gasparic, I. and Geraci, E. I. and Gernh{\"a}user, R. and Gnoffo, B. and G{\"o}bel, K. and {Gra{\~n}a-Gonz{\'a}lez}, A. and Haettner, E. and Hartig, A.-L. and Heil, M. and Heinz, A. and Hensel, T. and Holl, M. and Hornung, C. and Horvat, A. and Jedele, A. and Malenica, D. Jelavic and Jenegger, T. and Ji, L. and Johansson, H. T. and Jonson, B. and Jurado, B. and {Kalantar-Nayestanaki}, N. and Kazantseva, E. and {Kelic-Heil}, A. and Kiselev, O. A. and Klenze, P. and Kn{\"o}bel, R. and K{\"o}rper, D. and Kostyleva, D. and Kr{\"o}ll, T. and Kuzminchuk, N. and Laurent, B. and Lihtar, I. and Litvinov, Yu A. and L{\"o}her, B. and Martorana, N. S. and Mauss, B. and Morales, S. Murillo and M{\"u}cher, D. and Mukha, I. and Nacher, E. and Obertelli, A. and Pagano, E. V. and Panin, V. and Park, J. and Paschalis, S. and Petri, M. and Pietri, S. and Pirrone, S. and Politi, G. and Ponnath, L. and Revel, A. and Rhee, H.-B. and {Rodr{\'i}guez-S{\'a}nchez}, J. L. and Rose, L. and Rossi, D. and Roy, P. and Russotto, P. and Scheidenberger, C. and Scheit, H. and Simon, H. and {Storck-Dutine}, S. and Stott, A. and Sun, Y. L. and S{\"u}rder, C. and Tanaka, Y. K. and Taniuchi, R. and Tengblad, O. and Tisma, I. and T{\"o}rnqvist, H. T. and Trimarchi, M. and Velardita, S. and Vesic, J. and Voss, B. and Wamers, F. and Weick, H. and Wienholtz, F. and Zhao, J. and Zhukov, M.},
  year = 2025,
  journal = {Nature},
  pages = {1--6},
  publisher = {Nature Publishing Group},
  doi = {10.1038/s41586-025-08882-7},
  copyright = {2025 The Author(s), under exclusive licence to Springer Nature Limited},
  langid = {english}
}

@article{regnier2016fission,
  title = {Fission Fragment Charge and Mass Distributions in {\textsuperscript{239}}{{Pu}}(n,f) in the Adiabatic Nuclear Energy Density Functional Theory},
  author = {Regnier, D. and Dubray, N. and Schunck, N. and Verriere, M.},
  year = 2016,
  journal = {Phys. Rev. C},
  volume = {93},
  number = {5},
  pages = {054611},
  doi = {10.1103/PhysRevC.93.054611}
}

@article{regnier2019asymmetric,
  title = {From Asymmetric to Symmetric Fission in the Fermium Isotopes within the Time-Dependent Generator-Coordinate-Method Formalism},
  author = {Regnier, D. and Dubray, N. and Schunck, N.},
  year = 2019,
  journal = {Phys. Rev. C},
  volume = {99},
  number = {2},
  pages = {024611},
  doi = {10.1103/PhysRevC.99.024611}
}

@book{ring2004nuclear,
  title = {The {{Nuclear Many-Body Problem}}},
  author = {Ring, P. and Schuck, P.},
  year = 2004,
  publisher = {Springer Science \& Business Media},
  isbn = {978-3-540-21206-5}
}

@article{scamps2015superfluid,
  title = {Superfluid Dynamics of {\textsuperscript{258}}{{Fm}} Fission},
  author = {Scamps, Guillaume and Simenel, C{\'e}dric and Lacroix, Denis},
  year = 2015,
  journal = {Phys. Rev. C},
  volume = {92},
  number = {1},
  pages = {011602},
  doi = {10.1103/PhysRevC.92.011602}
}

@article{scamps2018impact,
  title = {Impact of Pear-Shaped Fission Fragments on Mass-Asymmetric Fission in Actinides},
  author = {Scamps, Guillaume and Simenel, C{\'e}dric},
  year = 2018,
  journal = {Nature},
  volume = {564},
  number = {7736},
  pages = {382},
  doi = {10.1038/s41586-018-0780-0},
  copyright = {2018 Springer Nature Limited},
  langid = {english}
}

@article{scamps2023spatial,
  title = {Spatial Orientation of the Fission Fragment Intrinsic Spins and Their Correlations},
  author = {Scamps, Guillaume and Abdurrahman, Ibrahim and Kafker, Matthew and Bulgac, Aurel and Stetcu, Ionel},
  year = 2023,
  journal = {Phys. Rev. C},
  volume = {108},
  number = {6},
  pages = {L061602},
  publisher = {American Physical Society},
  doi = {10.1103/PhysRevC.108.L061602}
}

@article{scamps2024quantal,
  title = {Quantal Effect on the Opening Angle Distribution between the Spins of the Fission Fragments},
  author = {Scamps, Guillaume},
  year = 2024,
  journal = {Phys. Rev. C},
  volume = {109},
  number = {1},
  pages = {L011602},
  publisher = {American Physical Society},
  doi = {10.1103/PhysRevC.109.L011602}
}

@misc{scamps2025uncertainty,
  title = {Uncertainty {{Principle}} and {{Angular Momentum Generation}} in {{Microscopic Fission Models}}},
  author = {Scamps, G. and Guilleux, A. and Regnier, D. and Bernard, A.},
  year = 2025,
  number = {arXiv:2512.02207},
  primaryclass = {nucl-th},
  publisher = {arXiv},
  doi = {10.48550/arXiv.2512.02207},
  archiveprefix = {arXiv}
}

@article{schunck2016microscopic,
  title = {Microscopic Theory of Nuclear Fission: A Review},
  author = {Schunck, N. and Robledo, L. M.},
  year = 2016,
  journal = {Rep. Prog. Phys.},
  volume = {79},
  number = {11},
  pages = {116301},
  doi = {10.1088/0034-4885/79/11/116301},
  langid = {english}
}

@article{schunck2022theory,
  title = {Theory of Nuclear Fission},
  author = {Schunck, Nicolas and Regnier, David},
  year = 2022,
  journal = {Progress in Particle and Nuclear Physics},
  pages = {103963},
  doi = {10.1016/j.ppnp.2022.103963},
  langid = {english}
}

@article{schunck2023microscopic,
  title = {Microscopic Calculation of Fission Product Yields for Odd-Mass Nuclei},
  author = {Schunck, N. and Verriere, M. and Potel Aguilar, G. and Malone, R. C. and Silano, J. A. and Ramirez, A. P. D. and Tonchev, A. P.},
  year = 2023,
  journal = {Phys. Rev. C},
  volume = {107},
  number = {4},
  pages = {044312},
  publisher = {American Physical Society},
  doi = {10.1103/PhysRevC.107.044312}
}

@article{simenel2010particle,
  title = {Particle {{Transfer Reactions}} with the {{Time-Dependent Hartree-Fock Theory Using}} a {{Particle Number Projection Technique}}},
  author = {Simenel, C{\'e}dric},
  year = 2010,
  journal = {Phys. Rev. Lett.},
  volume = {105},
  number = {19},
  doi = {10.1103/PhysRevLett.105.192701}
}

@article{simenel2014formation,
  title = {Formation and Dynamics of Fission Fragments},
  author = {Simenel, C. and Umar, A. S.},
  year = 2014,
  journal = {Phys. Rev. C},
  volume = {89},
  number = {3},
  pages = {031601},
  doi = {10.1103/PhysRevC.89.031601}
}

@article{tanimura2017microscopic,
  title = {Microscopic {{Phase-Space Exploration Modeling}} of {\textsuperscript{258}}{{Fm}} Spontaneous Fission},
  author = {Tanimura, Yusuke},
  year = 2017,
  journal = {Phys. Rev. Lett.},
  volume = {118},
  number = {15},
  doi = {10.1103/PhysRevLett.118.152501}
}

@article{verriere2021microscopic,
  title = {Microscopic Calculation of Fission Product Yields with Particle-Number Projection},
  author = {Verriere, Marc and Schunck, Nicolas and Regnier, David},
  year = 2021,
  journal = {Phys. Rev. C},
  volume = {103},
  number = {5},
  pages = {054602},
  publisher = {American Physical Society},
  doi = {10.1103/PhysRevC.103.054602}
}

@article{younes2009microscopic,
  title = {Microscopic Calculation of {{Pu240}} Scission with a Finite-Range Effective Force},
  author = {Younes, W. and Gogny, D.},
  year = 2009,
  journal = {Phys. Rev. C},
  volume = {80},
  number = {5},
  pages = {054313},
  doi = {10.1103/PhysRevC.80.054313}
}

@article{younes2011nuclear,
  title = {Nuclear {{Scission}} and {{Quantum Localization}}},
  author = {Younes, W. and Gogny, D.},
  year = 2011,
  journal = {Phys. Rev. Lett.},
  volume = {107},
  number = {13},
  pages = {132501},
  doi = {10.1103/PhysRevLett.107.132501}
}

@article{zhao2021microscopic,
  title = {Microscopic Self-Consistent Description of Induced Fission: {{Dynamical}} Pairing Degree of Freedom},
  author = {Zhao, Jie and Nik{\v s}i{\'c}, Tamara and Vretenar, Dario},
  year = 2021,
  journal = {Phys. Rev. C},
  volume = {104},
  number = {4},
  pages = {044612},
  publisher = {American Physical Society},
  doi = {10.1103/PhysRevC.104.044612}
}

@article{zhao2022timedependent,
  title = {Time-Dependent Generator Coordinate Method Study of Fission: {{Dissipation}} Effects},
  author = {Zhao, Jie and Nik{\v s}i{\'c}, Tamara and Vretenar, Dario},
  year = 2022,
  journal = {Phys. Rev. C},
  volume = {105},
  number = {5},
  pages = {054604},
  publisher = {American Physical Society},
  doi = {10.1103/PhysRevC.105.054604}
}

@article{marevic2021angular,
  title = {Angular Momentum of Fission Fragments from Microscopic Theory},
  author = {Marevi{\'c}, Petar and Schunck, Nicolas and Randrup, J{\o}rgen and Vogt, Ramona},
  year = 2021,
  journal = {Phys. Rev. C},
  volume = {104},
  number = {2},
  pages = {L021601},
  publisher = {American Physical Society},
  doi = {10.1103/PhysRevC.104.L021601}
}

@misc{regnier2026nucleoscope,
  title = {{{NucleoScope}}},
  author = {Regnier, David and Bernard, Alice and Newsome, Junah},
  year = 2026,
  url = {https://github.com/cea-phynu/nucleoscope}
}

\appendix

\section{Generation of wavefunctions: technical details}
\label{sec:wavefunction_details}

This section provides additional details on the way we computed the many-body wavefunctions 
used in this paper.
The Bogoliubov vacua considered for $^{20}$Ne and $^{252}$Cf all result from calculations
performed with the open-source HFB solver \texttt{HFB3}~\cite{dubray2025hfb3}.
The later assumes a time-reversal symmetry along with an axial symmetry of its solutions
and allows the parity breaking. 
We used here both the D1S and D1M Gogny EDF.
For $^{20}$Ne calculations, the quasiparticle states were expanded
onto a 1-center cylindrical harmonic oscillator basis with 13 major shells 
and the following parameters $b_z=b_{\perp}=1.6$ fm for the harmonic oscillator lengths along the z-axis and the perpendicular axes. We used a basis truncation scheme with $Q=1$.
For $^{252}$Cf studies we used a 2-centers harmonic oscillator basis with the parameters
$b_{\perp}=2.$, $b_z=2.65$ fm, a distance between the centers $d_0=18.5$ fm.
We retained 12 major shells and apply a truncation scheme with $Q=1.3$.

\begin{figure}[h!]
\includegraphics[width=1.0\linewidth]{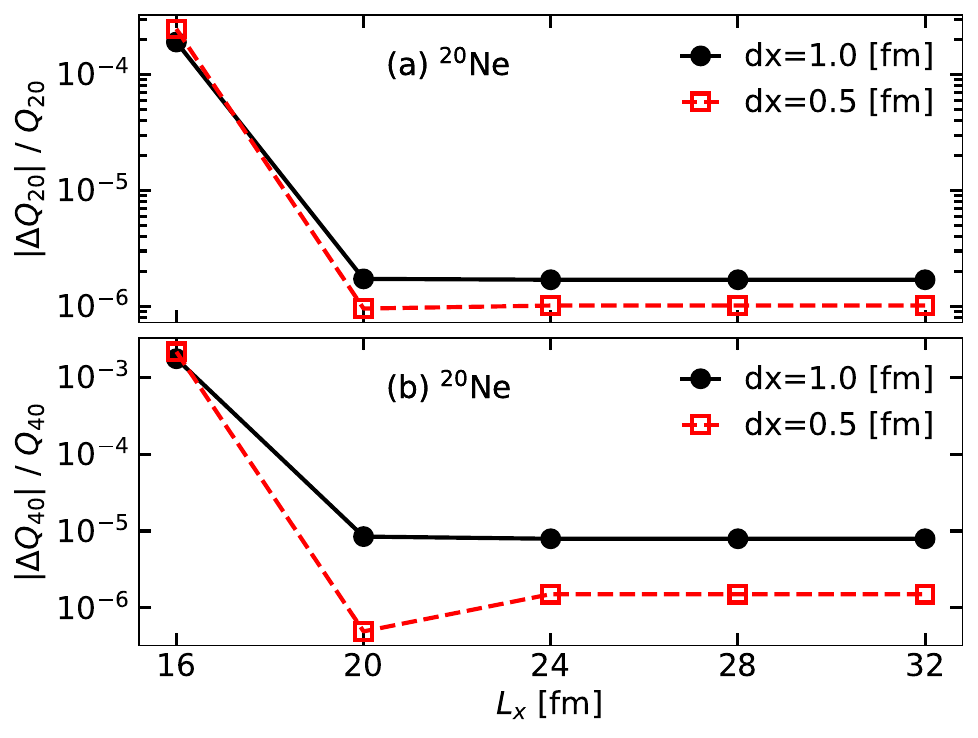}
\includegraphics[width=1.0\linewidth]{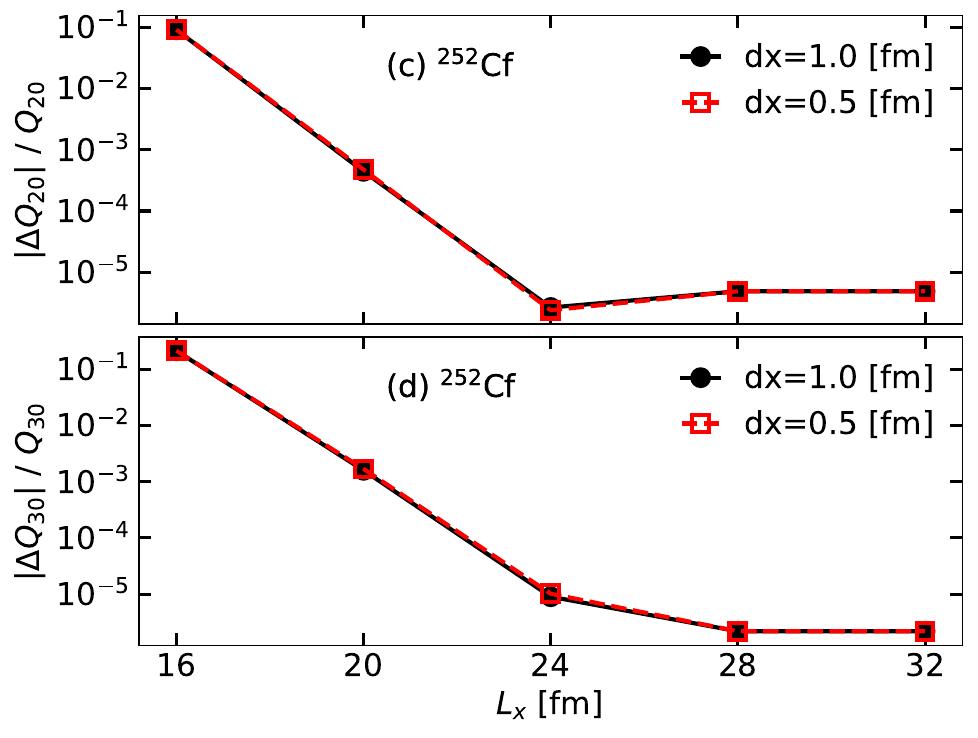}
\caption{Relative difference of multipole moments estimated on the mesh and on the original harmonic oscillator
from Bogoliubov vacua projected on good particle number.}
\label{fig:qlm_box_convergence}
\end{figure}

As mentioned in section~\ref{sec:method}, the Bogoliubov states are first expressed in their canonical representation and then discretized
on a regular mesh.
To choose the mesh length $L_x$ and the size of the cells $dx$, we compute the expectation values of the 
first two non-vanishing multipole moments in the projected HFB states considered in Sec.~\ref{sec:method}.
We perform these calculations both from the original states expressed in the harmonic oscillator basis and
from the one discretized on the mesh.
The later relies on a standard integration over the gauge angle to perform the particle number projection.
We used the Fomenko quadrature~\cite{fomenko1970structure} based on 20 discret angles.
We show the relative differences $\Delta Q_{lm} / Q_{lm}$ between the harmonic oscillator and mesh estimations in FIG.~\ref{fig:qlm_box_convergence}
as a function of the mesh parameters.
We see that the deterministic computations of these one-body observables in a box with a 
truncated basis is very well converged in terms of the box size and cell size for $L_x>24$ fm and $dx <1.$ fm.
At the chosen values of $dx=0.5$ fm and $L_x=32$ fm, the numerical bias for our calculation lies below $10^{-5}$.
The remaining inconsistency is likely coming from the canonical basis truncation explained in Sec.~\ref{sec:method}.

\section{Gelman-Rubin ratio to determine the burn-in period}
\label{sec:gelman-rubin}

To estimate the length of the burn-in period for one observable we rely on a statistical criteria called the Gelman-Rubin ratio
~\cite{brooks1998general,gelman1992inference}.
Looking at $n$ consecutive configurations sampled in $m$ independent Markov chains, we compute
estimators for the variance of the intra-chain means as well as the mean of 
the intra-chain variances.
\begin{align}
    B/n &= \frac{1}{m-1} \sum_{j=1}^m (\bar{O}_{j.} - \bar{O}{..})^2 \\
    W &= \frac{1}{m(n-1)} \sum_{j=1}^m \sum_{i=1}^n ( O_{ji} - \bar{O}_{j.})^2
    \label{eq:chap2:Wexpression}
\end{align}
Here $O_{ji}$ is the value of the observable kernel sampled at the $i$'th iteration 
of the $j$'th chain and we used the compact notations
\begin{align}
    \bar{O}_{j.} &= \frac{1}{n} \sum _{i=1}^n O_{ji} \, , \\
     \bar{O}_{..} &= \frac{1}{mn} \sum_{j=1}^m \sum _{i=1}^n O_{ji} \,  .
\end{align}
From this, we can build an estimator $\sigma^2$ of the actual variance of the pdf along with 
an estimator $V$ of the variance of the sample produced by the Markov chains:
\begin{align}
\sigma^2 &= \frac{n-1}{n}W + \frac{B}{n} \\
V &= \sigma^2 + \frac{B}{mn}.
\end{align}
The two first moment of the MCMC sample become representative of those of the actual pdf
when (i) the intra-chain variance $W$ stabilizes, (ii) the Gelman-Rubin ratio $R_{GL}=V/\sigma^2$ approaches one.

\section{Impact of $P_{flip}$ and $\sigma_{space}$ on the burn-in period}
\label{sec:parameter_impact_on_burnin}

\begin{figure}[h!]
    \centering
    \includegraphics[width=1.0\linewidth]{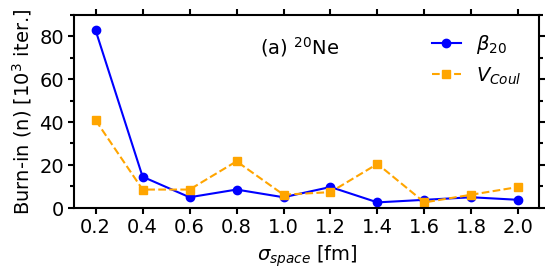}
    \includegraphics[width=1.0\linewidth]{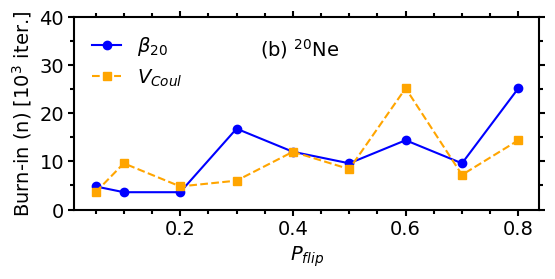}
    \includegraphics[width=1.0\linewidth]{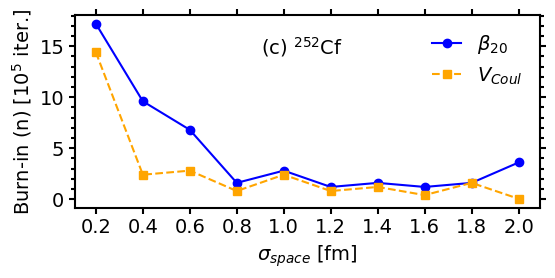}
    \includegraphics[width=1.0\linewidth]{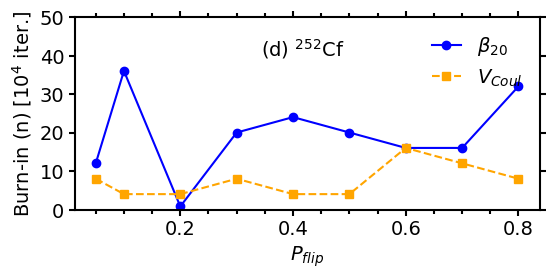}
    \caption{Burn-in periods as a function of the Markov chain parameters values $\sigma_{space}$ and $P_{flip}$. 
    The blue continuous line stands for the quadrupole moment and the yellow dashed line for the Coulomb interaction.
    The spatial width is varied while the spin flip probability is set to $P_{flip}=0.1$. The spin flip probability is varied while $\sigma_{space}=2$ fm.}
     \label{fig:burn-in-sigma}
\end{figure}

We performed a systematic study of the impact of the Markov chain parameters
$P_{flip}$ and $\sigma_{space}$ on the burn-in period as defined in \ref{sec:choice_mc_param}.
FIG.~\ref{fig:burn-in-sigma} shows the evolution of the the burn-in periods
obtained for a one- and two-body observables, in the case of $^{20}$Ne and $^{252}$Cf.
These results were obtained with 128 independent Markov chains going up to $n_{max}=240.10^3$ for $^{20}$Ne
and $8.10^{6}$ for $^{252}$Cf.

The burn-in does not show drastic variations with the Markov chain parameters. 
The burn-in period slightly increases with the spin-flip probability in the case of $^{20}$Ne while being quite insensitive to this parameter in an actinide.
We also show that the burn-in period decreases with $\sigma_{space}$ for both light and heavy nuclei.
It would be tempting to choose large $\sigma$ in our method but we found that the acceptance rate 
decreases nearly linearly with this parameter too.
Choosing $P_{flip}=0.1$ and $\sigma_{space}=1$ fm gives is the compromise we choose 
and gives an acceptance rate close to 0.5.

\end{document}